\documentclass{article} 

\usepackage{amsmath}
\usepackage{booktabs}
\usepackage{color}
\usepackage[usenames,dvipsnames]{xcolor}
\usepackage{float}
\usepackage{graphicx}

\usepackage[a4paper,bindingoffset=0.2in,%
left=1in,right=1in,top=1in,bottom=1in,%
footskip=.25in]{geometry}

\providecommand{\keywords}[1]{\textbf{\textit{Keywords:}} #1}

\usepackage{setspace}
\begin{document}
\title{Forecasting vegetation condition for drought early warning systems in pastoral communities in Kenya}
\maketitle

\begin{center}
\author{\textbf{Adam B. Barrett$^{a,b}$t, 
	Steven Duivenvoorden$^{a,c}$, 
	Edward E. Salakpi$^a$, 
	James M. Muthoka$^{d}$, 
	John Mwangi$^{e}$, 
	Seb Oliver$^{a,c}$ and 
	Pedram Rowhani$^{d}$*}\\
	$^{a}$ The Data Intensive Science Centre, Department of Physics and Astronomy, University of Sussex, Brighton BN1 9QH, UK\\
	$^{b}$ Sackler Centre for Consciousness Science, Department of Informatics, University of Sussex, Brighton BN1 9QJ, UK \\
	$^{c}$ Astronomy Centre, Department of Physics and Astronomy, University of Sussex, Brighton BN1 9QH, UK\\
	$^{d}$ School of Global Studies, Department of Geography, University of Sussex, Brighton, BN1 9QJ, UK \\
	$^{e}$ The National Drought Management Authority (NDMA), Lonrho House, Nairobi, Kenya \\
	*Corresponding author: P.Rowhani@sussex.ac.uk
	 }
\end{center}

\begin{abstract}

\noindent Droughts are a recurring hazard in sub-Saharan Africa, that can wreak huge socioeconomic costs. Acting early based on alerts provided by early warning systems (EWS) can potentially provide substantial mitigation, reducing the financial and human cost. However, existing EWS tend only to monitor current, rather than forecast future, environmental and socioeconomic indicators of drought, and hence are not always sufficiently timely to be effective in practice. Here we present a novel method for forecasting satellite-based indicators of vegetation condition. Specifically, we focused on the 3-month Vegetation Condition Index (VCI3M) over pastoral livelihood zones in Kenya, which is the indicator used by the Kenyan National Drought Management Authority (NDMA). Using data from MODIS and Landsat, we apply linear autoregression and Gaussian process modeling methods and demonstrate high forecasting skill several weeks ahead. As a bench mark we predicted the drought alert marker used by NDMA (VCI3M$<35$). Both of our models were able to predict this alert marker four weeks ahead with a hit rate of around 89\% and a false alarm rate of around 4\%, or 81\% and 6\% respectively six weeks ahead. The methods developed here can thus identify a deteriorating vegetation condition well and sufficiently in advance to help disaster risk managers act early to support vulnerable communities and limit the impact of a drought hazard.
\end{abstract}

\section{Introduction} 

Droughts are a major threat globally as they can cause substantial damage to society, especially in regions that depend on rain-fed agriculture. They particularly impact food security by significantly reducing agricultural production \cite{Lesk2016} and raising food prices \cite{Nelson3274,BROWN201531}, which often leads to increased levels of malnutrition, migration, disease, and other health concerns \cite{10.1093/rsq/hdr006,Stanke}. The majority of droughts occur in sub-Saharan Africa \cite{emdat} where many communities rely on predictable rainfall patterns for their livelihood.

In East Africa, the main economic activity in the arid and semi-arid lands (ASAL) is subsistence rain-fed agriculture, as well as livestock farming using pastures and grasslands as the main source of fodder. As a result, the pastoral and agro-pastoral communities who live in these drylands are particularly vulnerable to drought \cite{Nyong2007,orindi2007}, especially since their existing coping strategies have been compromised by population growth and land use change in recent years \cite{Galvin2001}. Governments and donor agencies in the region have thus developed several tools and early warning systems (EWS) to mitigate the impact of droughts on pastoralists.

Most EWS tend to monitor current key biophysical and socio-economic factors to assess the possible exposure of vulnerable people to specific hazards. However, once the impacts are visible, it may be too late to mitigate the consequences \cite{kogan}. Hence there is growing interest in moving toward a proactive humanitarian approach to disasters by developing preparedness actions based on climate forecasts \cite{CoughlandePerez2015,Lopez2018, wilkinson2018forecasting}. Additionally, it is estimated that being better prepared before a drought hits significantly reduces the costs and losses from these disasters \cite{Venton2012}. Hence, EWS now increasingly include expert knowledge and qualitative assessments of seasonal climate forecasts to assess the future development of food security, and define actions to mitigate possible losses \cite{CoughlandePerez2015,TozierdelaPoterie2015}. However for drought conditions, a meteorological drought does not always lead to negative agricultural outputs \cite{BHUIYAN2006289}. There is thus a growing interest to include forecasts of the impacts of these hazards \cite{wmo2015wmo,nhess-2018-26,Sutanto2019}.

In Kenya, following several periods of intense drought, the government established the National Drought Management Authority (NDMA) in 2016, to set up and operate a drought EWS, as well as to establish drought preparedness strategies and contingency plans. The NDMA provides monthly bulletins assessing food security in the 23 ASAL regions using current  biophysical (e.g., rainfall, vegetation condition) and socio-economic (production, access, and utilisation) factors. One key biophysical indicator used by the NDMA drought phase classification is based on the Vegetation Condition Index(VCI) \cite{KOGAN199591,rs8040267,RULINDA201132,ROJAS2011343}. 

The VCI, which expresses the Normalized Difference Vegetation Index (NDVI) in terms of where it currently lies within its expected range for the given pixel, is one of a number of satellite-based indicators that have been developed to detect and monitor drought \cite{zargar2011review}. While there is little agreement between VCI and precipitation-based meteorological drought indicators \cite{quiring2010evaluating,BHUIYAN2006289}, it is strongly linked to agricultural production and widely used to identify drought onset, intensity, duration, and impact \cite{rs8030224}. The NDMA uses the 3-month averaged VCI (VCI3M) in its operational EWS \cite{rs8040267}. Once the VCI3M goes below a threshold of 35, the NDMA triggers a rapid food security assessment and has access to the National Drought Contingency Fund in order to implement its preparedness strategies and contingency plans.

The main goal of this paper is to explore machine-learning techniques to forecast the vegetation indices that are commonly used in the pastoral areas of Kenya to monitor droughts. In order to provide useful information to drought risk managers, we aim to identify the right balance between forecast lead time and uncertainty. To this end, we evaluated the performance of our approaches up to ten weeks ahead.

Based on NDMA's experience, we particularly focused on the pastoral livelihood zones as the VCI3M is more reliable in identifying drought condition for grazing and browsing in the more arid regions of the country.  Several studies have developed statistical and machine-learning approaches \cite{Udelhoven,meroni2014early,Zambrano:2018,vrieling2016early} to predict end-of-season crop, forage and biomass production. Recently, \cite{matere2019predictive} developed a decision support tool based on a mechanistic model to estimate 6-monthly forecasts of forage condition. Here, we specifically focus on Gaussian Process (GP) modelling \cite{gpm}, and linear autoregressive (AR) modelling \cite{Hamilton94} to forecast NDVI and VCI3M, which are derived from both Landsat (every 16 days at 30\,m resolution) and the MODerate resolution Imaging Spectroradiometer (MODIS - daily data at 500\,m resolution). GP modelling uses kernel-based non-parametric Bayesian inference on the structure of correlations between observations, and is widely applied to classification, interpolation, change detection and forecasting problems \cite{955315,Chandola2010SCALABLETS,7487896,rs11050481}. Linear AR is the regression of future observations on past observations, assuming a linear dependence. Previously it has been performed on monthly (i.e.~temporally more sparse) NDVI data, see for example \cite{Asoka:2015} and \cite{Papagiannopoulou}, with mixed results in terms of forecasting potential ($R^2$-scores between 0 and 0.4 at a lead time of one month). 

\section{Study area}
In Kenya, the livestock sector accounts for 13\% of the national GDP and 43\% of its agricultural GDP. Livestock farming mainly occurs in the ASAL which cover about 80\% of the country \cite{UNDP2013,FAO2014}. In these regions, the pastoral communities rely on pastures and grasslands as the main source of fodder \cite{Behnke2011}. Thus, providing information on pasture productivity to these communities is key in times of drought.

\begin{figure}[!h]
	\centering
	\includegraphics[trim = 0mm 0mm 0mm 10mm,width=13.5 cm]{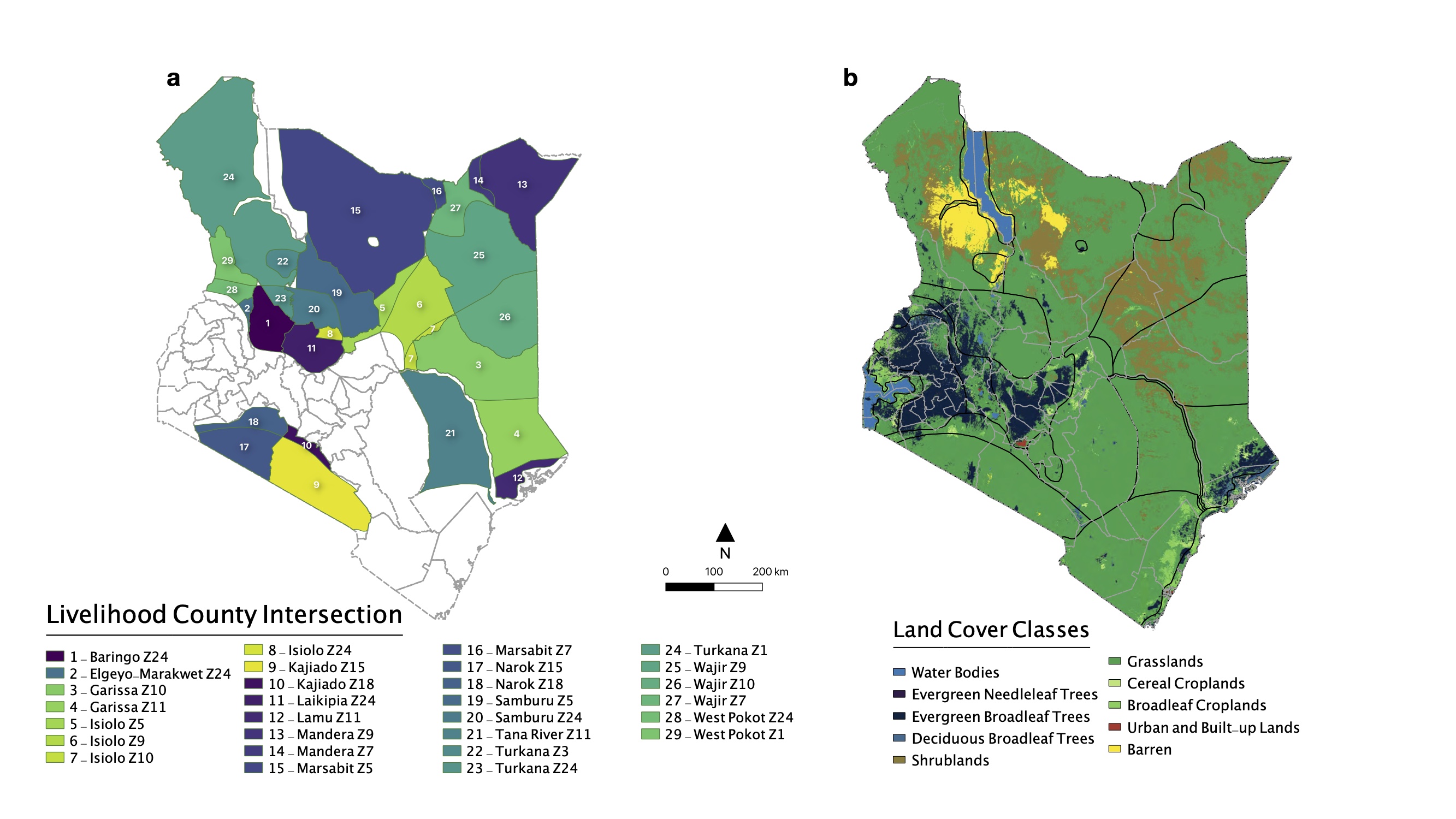} 
	\caption{Maps of Kenya showing (a) Livelihood Zones and County intersections (Regions of Interest (ROI)) from which pixels were sampled for analysis, and (b) land-cover classification (according to the MODIS MCD12Q1 data). Analyses were performed for 29 regions, defined by pastoral livelihood zone and county intersections. A map showing the livelihood zones can be found in Fig. \ref{fig:l_zone} in the Supplementary Material.} \label{fig:l_zone_roi}
\end{figure}

For the ASAL regions, the NDMA reports every month the VCI3M value at county level as well as over the different livelihood zones within the county. This study focused on the 10 (agro)-pastoral livelihood zones (see Fig.~\ref{fig:l_zone}), which cross 15 counties. The names of the 29 livelihood zone county intersections can be found in Fig.~\ref{fig:l_zone_roi}; these are our regions-of-interest, which we refer to simply as `regions'.

\newpage

\begin{figure} 
	\centering
	\includegraphics[trim = 0mm 0mm 0mm 20mm,width=11.5 cm]{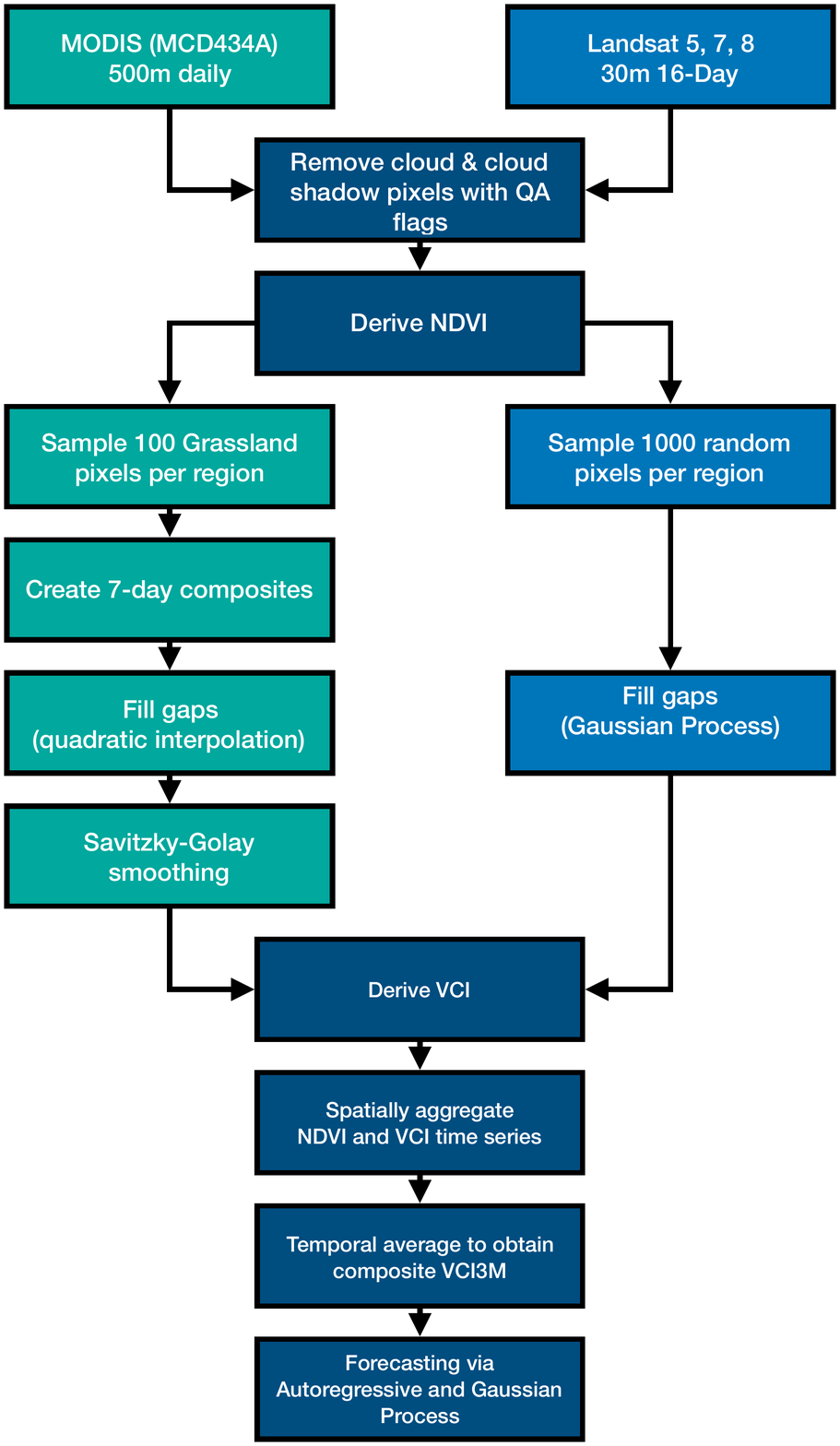}
	\caption{A flow chart of the data processing and analysis.} \label{fig:flowchart}
\end{figure}
\newpage

\section{Methods} \label{sec:methods}
This research is based on two satellite-based Earth observation datasets, Landsat and MODIS. Description and justification of data selection, and a comparison between the two datasets can be found in \ref{sec:datasets}. It should be noted that the analysis is based on a random subsample of the pixels within each of the 29 regions (Fig.~\ref{fig:l_zone_roi}). A summary of the entire work from data preparation to forecasting drought can be seen in Fig.~\ref{fig:flowchart}.

\subsection{Data preprocessing} \label{sec:preprocessing}

\subsubsection{Landsat}
\begin{figure}
\label{review:GP_sketch}
    \centering
    \includegraphics[width=10cm]{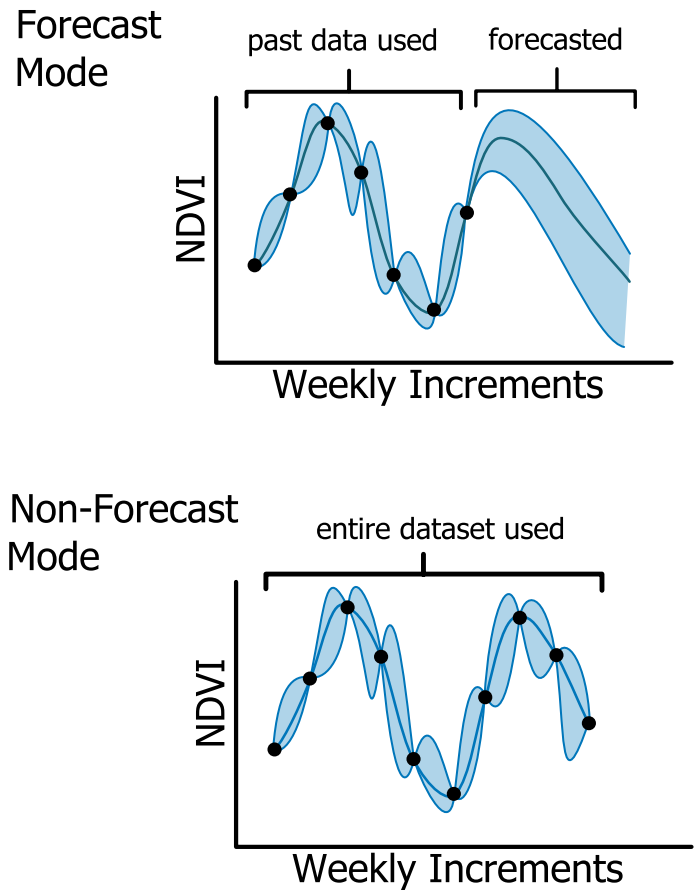}
    \caption{Illustration of the GP approach used for the Landsat data. In ``forecast mode", the correlations in the data up to a given date furnish a GP model, which can then be used for forecasting. In the ``non-forecast mode", the entire time series is used to train the GP, and provide a ground truth for the forecast.}
    \label{fig:GP_illustration}
\end{figure}

Temporal gridding and gap-filling on the Landsat data was done using Gaussian Process (GP) regression. For a given pixel, the GP regression took raw data as input, fit a temporal correlation structure to the data, and used this to output a time series of expected NDVI values, with observations provided every Saturday over the studied time period; see Figure~\ref{fig:GP_illustration} for an illustration and \ref{sec:GPexplain} for details. Two versions of GP gap-filling were carried out, which we refer to as forecast mode and non-forecast mode. For the non-forecast mode, the full time series from the given pixel were used to train the GP. The non-forecasting mode was used as the ``ground truth'' to test forecasts against. The forecast mode, by contrast, only used data up to a certain date, whichever date a forecast was being attempted from - since when doing forecasting with a near real-time data stream, one does not have access to future data.

\subsubsection{MODIS}

Weekly NDVI composites were obtained for each pixel by taking the mean (after cloud masking) of all available data over a 7-day time period. Gaps in the weekly time series were then filled using quadratic interpolation. Gaps longer than 6 weeks were left unfilled, see \ref{sec:Lmax} for  details. The gap-filled time series were then smoothed using the Savitzky-Golay method \cite{Savitzky1964} to filter high-frequency measurement noise. The smoothing involved fitting, for each pixel, a polynomial to a window centred on the observation, and then replacing that observation with the output of the polynomial fit. The polynomial order was set to 2 (i.e.~quadratic function) and the window length to 7 weeks. (Note that the combined interpolation and smoothing procedure does two rounds of quadratic interpolation where there are gaps, but that these are distinct: the interpolation fills a gap of up to 6 weeks with one quadratic function, while the smoothing modifies only one observation per fitted quadratic function.)

\subsection{Indices}
On both datasets, VCI time series were constructed from the NDVI time series according to the formula:
\begin{equation}
\rm{VCI}_i = 100 \times \frac{\rm{NDVI}_i - \rm{NDVI}_{min,i}}{ \rm{NDVI}_{max,i} - \rm{NDVI}_{min,i}},
\label{vci_eq}
\end{equation}
where $\rm{NDVI}_{min,i}$ and $\rm{NDVI}_{max,i}$ are the minimum and maximum observed values for the NDVI of the pixel for the week of the year at time point $i$. The data within each region were aggregated taking the mean of the sampled pixels at each time point. Thus forecasting was applied on a single time series for each region. Finally, VCI3M was calculated as the mean VCI across the 12 weeks leading up to the given time point. Additionally, aggregate time series of NDVI anomaly were constructed (i.e., seasonal mean-subtracted NDVI, sometimes referred to as absolute anomaly; results for forecasting this can be found in \ref{sec:further_results}).

With Landsat data, the mean, maximum and minimum value for the NDVI in \eqref{vci_eq} was computed using the non-forecast mode GP interpolated time series. Then forecast mode and non-forecast mode versions of each index were created. With the MODIS data, since large gaps were unfilled, whenever there were fewer than 25 individual pixel observations from a particular region at a given time, it was decided that there should be no datum in the aggregate VCI time series for that region (i.e., there should be a gap in the time series). Additionally, if the current aggregate NDVI observation was not present, a gap was placed in the VCI3M time series. Else, the mean was taken over all present observations from the most recent 12 weeks.

\subsection{Forecasting}\label{forecast_method}
Machine-learning techniques offer a data-driven, empirical route to forecasting. Many different data inputs could be used to forecast these vegetation indices (e.g. precipitation and precipitation forecasts). However, perhaps the most simple is to use the past history of the indices themselves. This has the practical benefits of readily available data over large areas. Additionally, this approach will also take advantage from the fact that these indices are subject to plant growth and climate cycles giving periodic behaviour on large temporal scales that can be empirically modelled, while external perturbations, such as water availability, have persistent impact providing correlations on short temporal scales. Forecasts of NDVI anomaly and VCI3M were made using two separate methods, respectively based on Gaussian Process modelling (GP) and linear autoregressive (AR) modelling.

GP forecasting was performed by fitting a GP to the forecast mode aggregate time series for the index in question, and then using the GP to extrapolate. For details on GP modelling, see \ref{sec:GPexplain}. The key step involved fitting a temporal correlation structure to the time series, i.e.~a kernel $k(t,t^{\prime})$ that describes the covariance between the index at any two times $t$ and $t^{\prime}$. The kernel with the highest evidence was the Radial Basis Function (RBF):

\begin{align}
	k_{\rm{RBF}}(t,t') &= \sigma_{\rm{RBF}}^2 \exp{\Big( -0.5 \frac{|t-t'|^2}{l_{\rm{RBF}}^2} \Big) }\,,
\end{align}
where $\sigma_{\rm{RBF}}^2$ and $l_{\rm{RBF}}$ are the signal variance and the length scale, respectively, and the modelling was carried out with the best fit version of this.

AR forecasting was performed with the following model-fitting and extrapolation method. For forecasting $n$ weeks ahead, the following model was fit:

\begin{equation}
X_{t+n}=\sum_{i=0}^{p-1}a_iX_{t-i}+\epsilon_t\,, \label{eq:AR1}
\end{equation}
where $X$ is the index in question, subscripts denote the date (week), $a_i$ are model coefficients, $\epsilon_t$ are the residuals (i.e.~the errors), and $p$ is called the model order. (This model assumes zero mean, so for VCI3M, the mean was removed prior to fitting the model, and then added back again after using the model to forecast the deviation from the mean.) Fitting the model to a segment of data involved finding the model coefficients that gave the minimum sum-square error, i.e.~led to residuals with the minimum variance. To make a forecast, the model was fit using the most recent $T$ consecutive observations (where $T$ is called the training segment length), and then used to predict the observation $n$ weeks after the most recent observation. This forecasting method was carried out along the entire available time series, fitting a distinct model to each segment of length $T$. The model order was set to $p=3$ and the training segment length was set to $T=200$, since forecast skill plateaued at these values.

\subsection{Forecast assessment}
Several metrics were used to assess the performance of the forecast methods tested on the data. In addition to RMSE, the $R^2$-score and the percentage of standard deviation remaining, $S$, were used. These are given by:

\begin{align}
	R^2{\text -}{\rm score} &= 1 - \frac{\sum_i (y_i - f_i)^2 }{\sum_i (y_i - \bar{y})^2}, \label{eq:r2} \\
	S &= 100 \times \frac{\sqrt{ \sum_i (y_i-f_i)^2 }}{\sqrt{  \sum_i (y_i-\bar{y})^2 }} \label{eq:S}\,,
\end{align}
where the $y_i$ are the true data, and the $f_i$ are the forecasts. Note that $S\equiv 100\times\sqrt{1-R^2{\text -}\rm{score}}$. To test for bias, we performed a linear regression of the actual index on the forecast index, and extracted the slope and intercept. Finally, receiver operating characteristic (ROC) curves were constructed for forecast-based drought-alert detection. 

These performance indicators were also used to assess the sensitivity of our methods in space (comparing the results by region) and in time (to account for seasonality). Additionally, the forecast methods were evaluated for various drought categories \cite{rs8040267}, compared against a persistence forecast (i.e.~forecast obtained by taking the most recent observation to be the forecast value), and the impact of data gaps on forecast performance was analysed.

Forecasts on the MODIS data were assessed from January 1st 2004 onward, which was approximately the earliest date for which there were sufficient prior data for the AR method to be applied. Forecasts on the Landsat data were assessed only from January 1st 2014 onward, since the GP gap-filling method required training on data up to this date.

\section{Results} \label{sec:results}

\subsection{Forecast value accuracy}

\begin{figure}
	\centering
	\includegraphics[trim = 30mm 35mm 0mm 10mm,width=5.4 cm]{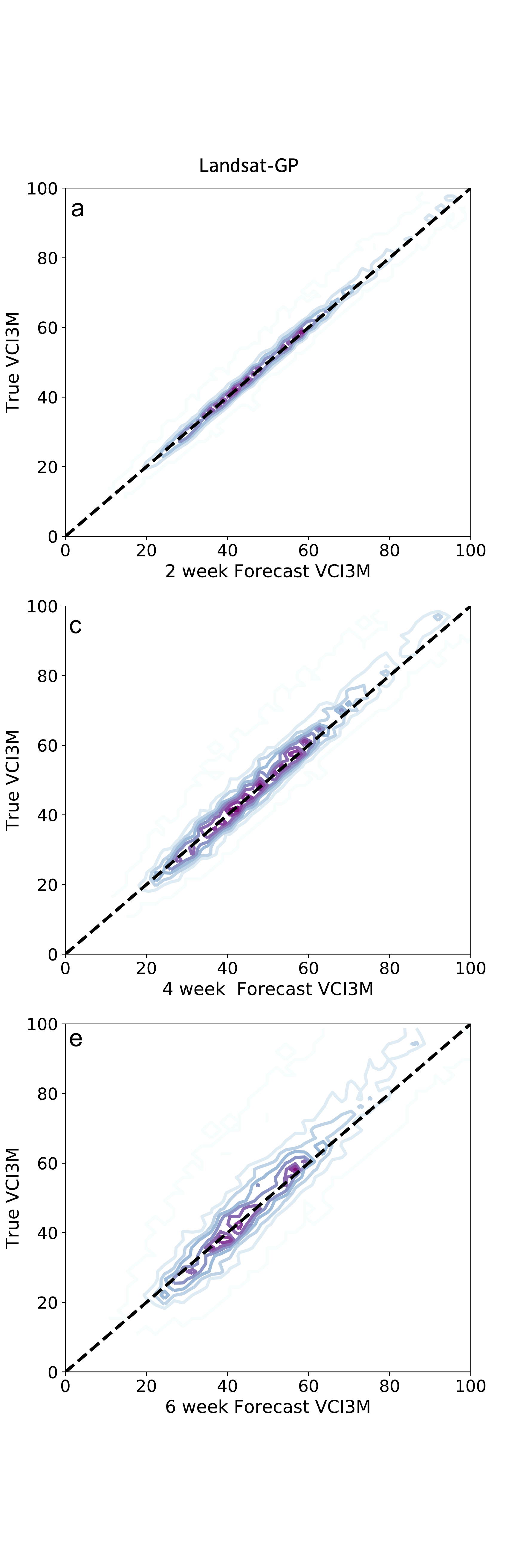} \qquad \includegraphics[trim = 30mm 35mm 0mm 10mm,width=5.4 cm]{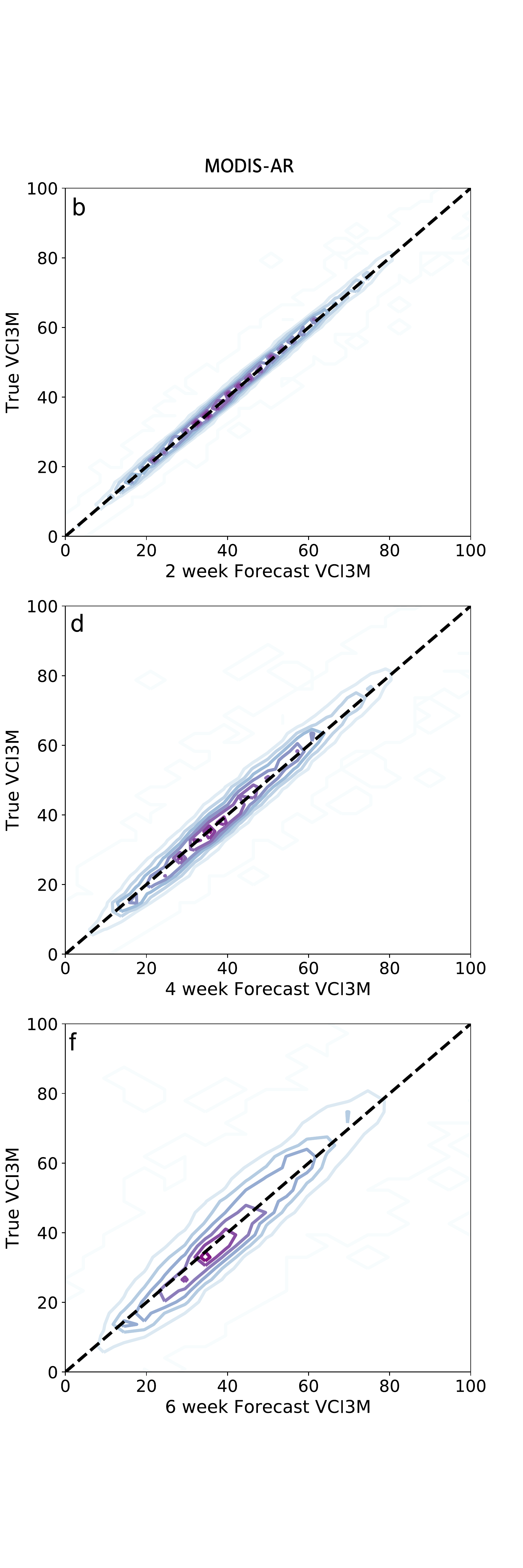}
	\caption{Contour plots of VCI3M against two, four and six weeks VCI3M forecasts. (a,c,e) show forecast performance for the GP method on Landsat data, and (b,d,f) show forecast performance for the AR method on MODIS data (across the 19 regions for which a 4 week forecast was possible more than 50\% of the time, see main text for details).} \label{fig:gpcontour}
\end{figure}

\begin{table}
	\small
	\caption{ Performance statistics of VCI3M forecasts with lead times of 2, 4 and 6 weeks. Data for slope and intercept show ordinary least squares estimates $\pm$ standard error.} \label{tab:stats}
	\centering
	\begin{tabular}{l|ccc|ccc} 
		\toprule
		& & \textbf{Landsat GP} & & &\textbf{MODIS AR} \\
		& \textbf{2} & \textbf{4} & \textbf{6} & \textbf{2} & \textbf{4} & \textbf{6} \\
		& \multicolumn{3}{c|}{\textbf{weeks }}& \multicolumn{3}{c}{\textbf{weeks }}\\

		\midrule

		$R^2$-score  &0.99  &0.95  &0.87  &0.99& 0.96& 0.88\\
		RMSE &1.8  &4.2  &6.8  &1.8& 4.3& 7.0 \\
		slope &1.0$\pm0.0$  &1.1$\pm0.0$ &1.1$\pm0.0$ &1.0$\pm 0.0$ &1.0$\pm0.0$& 1.0$\pm 0.0$ \\
		intercept &-1$\pm0$  &-2$\pm0$ &-3$\pm0$&0$\pm0$& 0$\pm0$& 0$\pm0$\\
		\bottomrule
	\end{tabular}
\end{table}

The GP and AR forecasting methods were applied, on each of the two datasets, to regional aggregate VCI3M time series. We focus on performance results of GP forecasting on Landsat data and AR forecasting on MODIS data since these two combinations of data and forecasting method performed the best (as measured by $R^2$-score). We looked at lead times of up to ten weeks (see Figures \ref{fig:GP_NDVI_forecast} and \ref{fig:NDVI_forecast}). However, due to increasing uncertainty, the results provided here focus on two to six weeks forecasts of VCI3M.

Contour plots of forecast against actual data for two, four and six week forecasts are shown in Fig.~\ref{fig:gpcontour}. Table \ref{tab:stats} shows the $R^2$-scores, RMSE, slope and intercept from each of these plots, and demonstrates that there is substantial forecast skill from each method at each lead time ($R^2$-scores are substantial), and that the forecasts are unbiased (slopes are all approximately 1, and intercepts approximately 0). Corresponding results for NDVI anomaly time series can be found in the Supplementary Material, in Fig.~\ref{fig:contourNDVI} and Table \ref{tab:stats2}. The much higher $R^2$-scores for VCI3M compared to NDVI anomaly is due to the fact that VCI3M is the 12 week composite of weekly VCI observations, and thus there is much greater correlation between VCI3M at time $t$ and VCI3M at time $t+(2,$ 4, 6) weeks due to overlap in the composited weeks. Notwithstanding this, each forecasting method performed substantially better than a persistence forecast of VCI3M (i.e.~taking the most recent observation to be the forecast value). For example, the AR method on the MODIS data achieved an RMSE of approximately half that of the persistence forecast for a lead time of 4 weeks, see Fig.~\ref{fig:persistence} in the Supplementary Material.

\begin{figure}
	\centering
	\includegraphics[trim = 50mm 0mm 0mm 0mm,width=10cm]{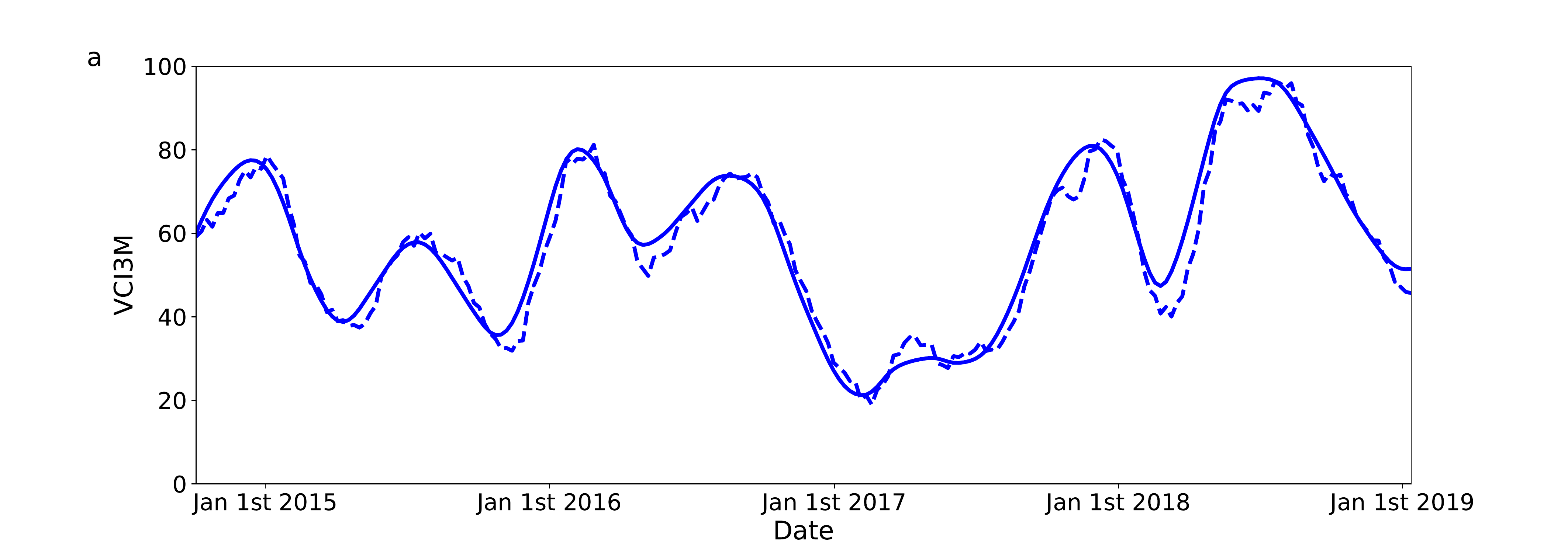}
    \includegraphics[trim = 50mm 0mm 0mm 0mm,width=10cm]{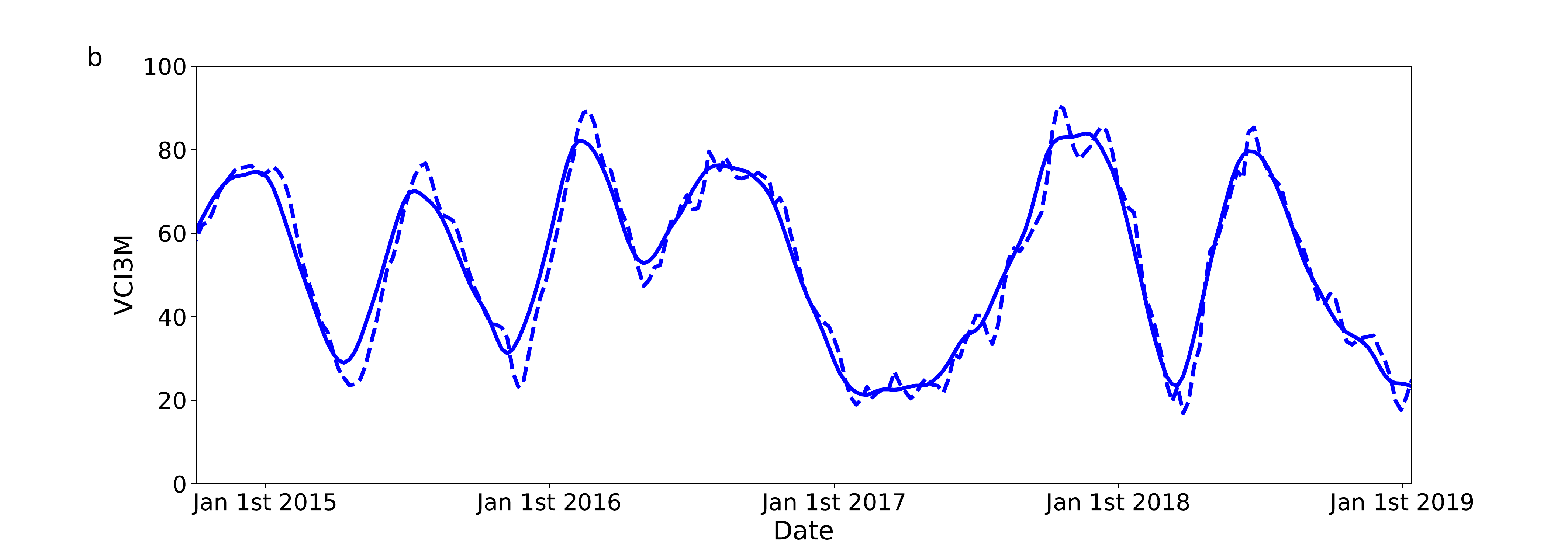}
	\caption{Sample aggregate VCI3M} time series from the intersection of Baringo county and livelihood zone 24 from (a) Landsat and (b) MODIS (solid lines). Dotted lines show forecasts at a lead time of 4 weeks, as given by the GP method on the Landsat data, and the AR method on the MODIS data. \label{fig:ndvi_lk}
\end{figure}

For both methods, the forecast time series sometimes lag behind the true time series, since changes not foreseen by the models are incorporated only once they are observed, see Fig.~\ref{fig:ndvi_lk} for examples. 
The two methods sometimes make different types of error. The GP method is more likely to predict a value that is closer to the long-term mean than the true value. This is because \textit{a priori} to taking into account the most recent observations, the model assumes the forecast observation will be equal to the long term mean. By contrast, the AR method is more likely to predict a continuation of the recent trend. This is because the model assumes a continuation of the recent frequency profile, so if a faster-than-average trend is seen in either direction, the trend will be predicted to continue \cite{Hamilton94}.

Due to the presence of non-interpolated gaps in the MODIS time series, there were weeks when a forecast assessment was not carried out on these data, see Table \ref{tab:IL choices} for details.  For 15 of the regions, a 4 week forecast could be made on more than 90 percent of weeks; however, for some regions, a forecast could rarely be made, see Fig.~\ref{fig:R2plots} in the Supplementary Material.

\begin{figure}
	\centering
	\includegraphics[trim = 30mm 15mm 0mm 10mm,width=5.4 cm]{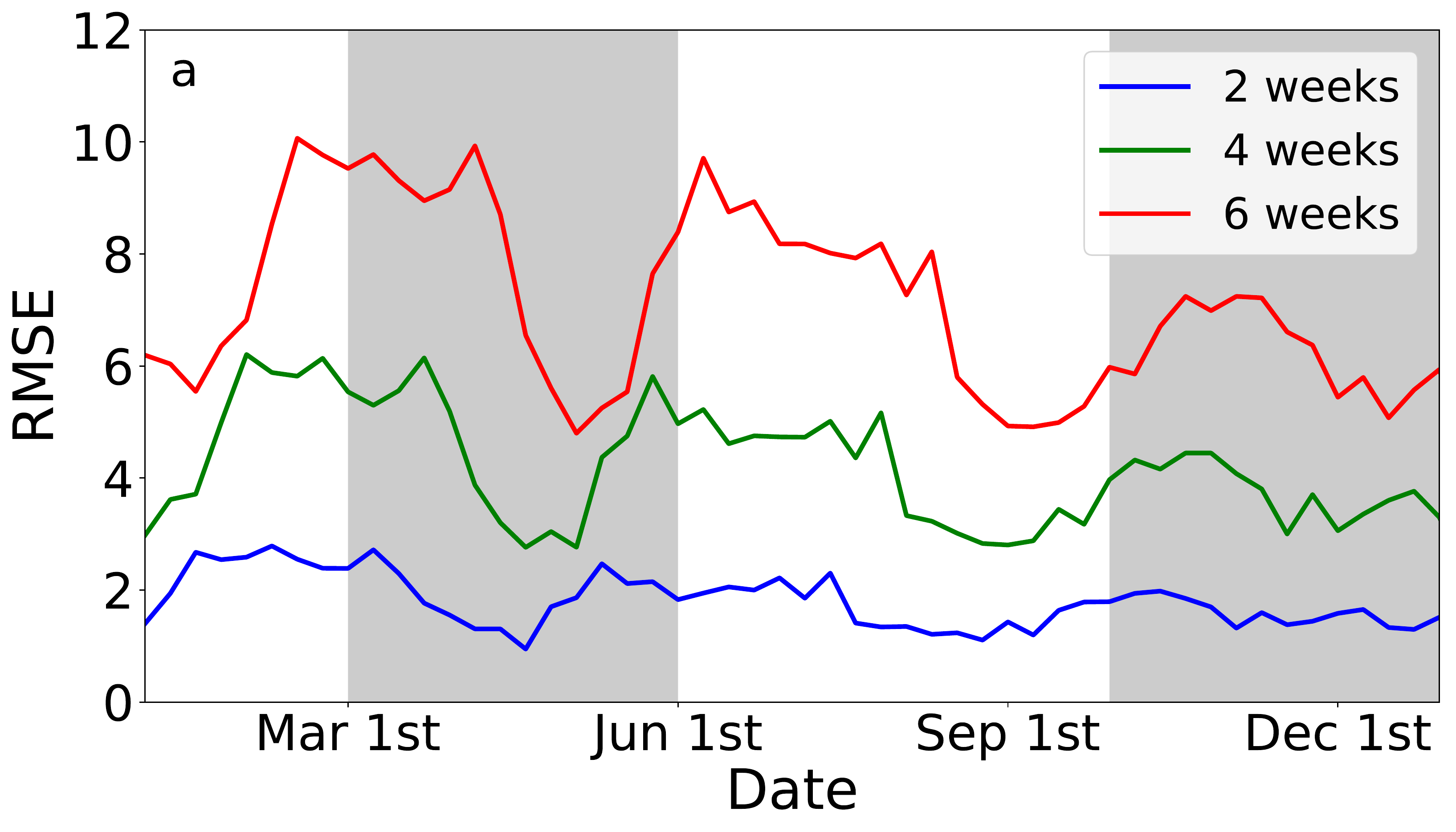} \qquad \includegraphics[trim = 30mm 15mm 0mm 10mm,width=5.4 cm]{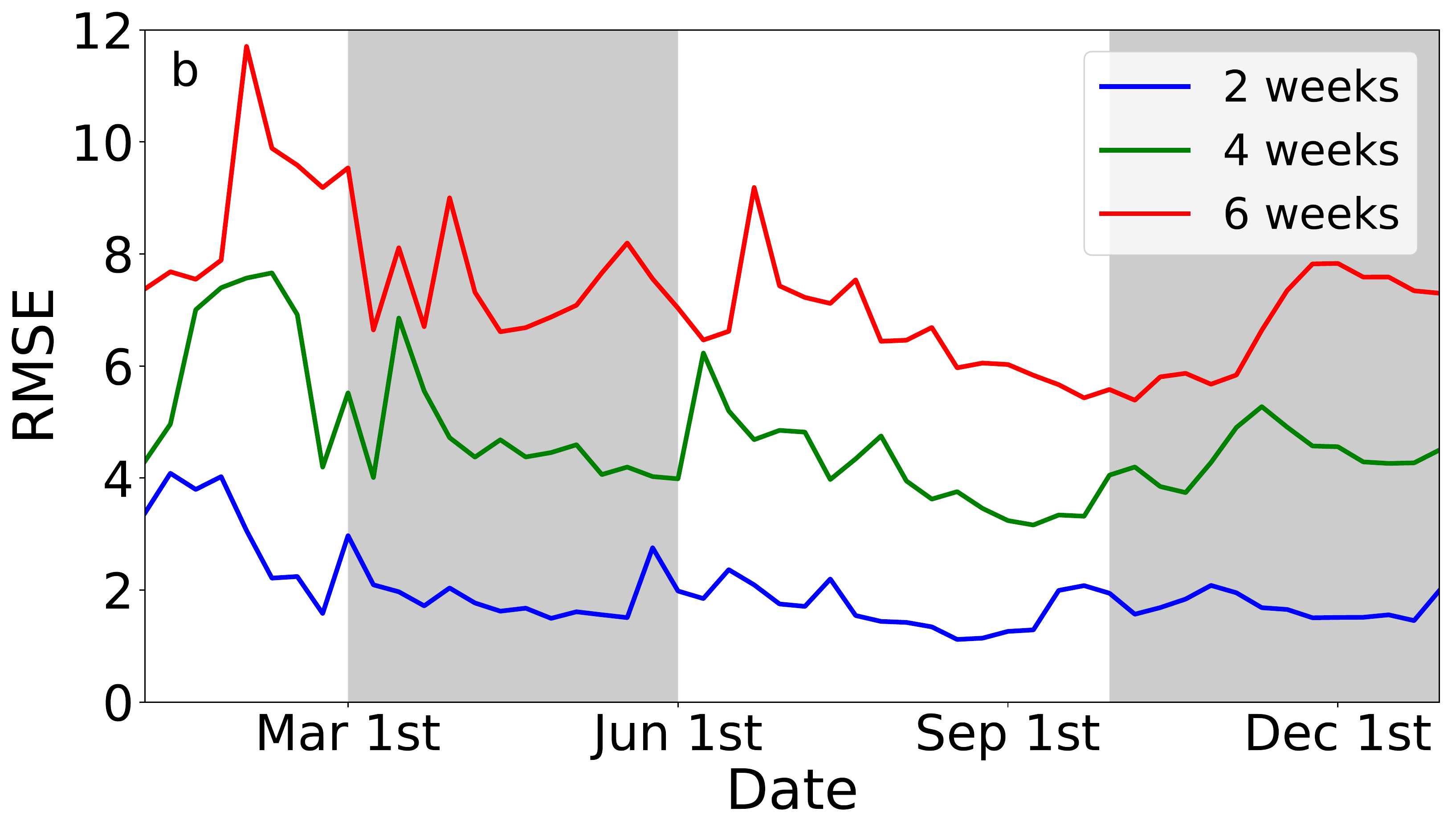}
	\vspace{0.5cm}
	\caption{RMSE of VCI3M forecast for each week of the year. (a) GP forecasting on Landsat data. (b) AR forecasting on MODIS data. Grey shading indicate the rainy seasons, March-May and October-December.} \label{fig:seasonal}
\end{figure}

Additional checks were included to test the sensitivity of the methods to drought severity and seasonality. 
The methods, when computed separately for each of the five categories on the NDMA drought scale \cite{rs8040267}, perform better in terms of absolute RMSE when there was a state of drought than when the vegetation condition was normal or wet (Table \ref{tab:RMScategories}). This could be explained by the fact that when conditions are relatively normal, the subsequent conditions could go in various directions. However, when there is an extreme drought, it is likely that it will persist (because vegetation cover is already below-normal). Note though that the relative RMSE as a proportion of the VCI3M value appears to be similar during drought than during normal vegetation condition. 
Neither method exhibited large seasonal differences, see Fig.~\ref{fig:seasonal}. However, for the GP method applied to Landsat data at a 6-week lead time, RMSE is high at the start of the season but drops noticeably during the course of the season, suggesting that it is harder to predict how a season will start but  easier to forecast how it will proceed once started.

\begin{table}
	\small
	\caption{RMSE in VCI3M forecast, for the true vegetation condition belonging to the different categories of drought, at lead times of 2, 4 and 6 weeks. Drought categories are defined by the VCI3M index: wet by VCI3M$>$50;  normal by 35$<$VCI3M$<$50; moderate drought by 20$<$VCI3M$<$35; severe drought by 10$<$VCI3M$<$20; and extreme drought by VCI3M$<$10.
	(The extreme drought criterion was not met in any of the Landsat data.)} \label{tab:RMScategories}
	\centering
	\begin{tabular}{l|ccc|ccc} 
		\toprule
		\textbf{Drought category} & & \textbf{Landsat GP} & & &\textbf{MODIS AR} \\
		& \textbf{2 } & \textbf{4 } & \textbf{6 } & \textbf{2 } & \textbf{4 } & \textbf{6}\\
		& \multicolumn{3}{c|}{\textbf{weeks}} &  \multicolumn{3}{c}{\textbf{weeks}}\\
		\midrule
		Wet  & 2.2 & 5.3 & 9.0 &2.2& 4.8& 7.5\\
		Normal & 1.7 & 3.4 & 5.0 &1.6& 4.0& 6.5 \\
		Moderate drought & 1.5 &3.2 & 5.0&1.5&3.7& 5.7\\
		Severe drought, & 1.1 &2.5 &5.5&1.4& 3.3& 5.4\\
		Extreme drought & &&& 1.1 & 2.9   & 4.8         \\
		\bottomrule
	\end{tabular}
\end{table}

The fact that seasonal differences in RMSE are not substantial provides reassurance that the forecast accuracy estimates are not inflated by the gap-filling during preprocessing. If this were the case, there would be a sustained drop in RMSE during the more overcast months of the year (March to May and October to December). While the GP forecasts on the Landsat data were computed from time series on which no future data were used for the interpolations (see \ref{sec:preprocessing}), for the MODIS data interpolations did make use of future data. Therefore, to obtain further reassurance that performance estimates of AR forecasting on the MODIS data were not inflated, a plot was made of RMSE at 4 weeks lead time against percentage of pixels from which a good observation was obtained on the date of the forecast, see Fig.~\ref{fig:percentclear} in the Supplementary Material. There was no apparent correlation (Pearson coefficient was 0.01), and hence it was concluded that the gap-filling was not leading to inflated forecast performance.

\subsection{Drought event forecast: ROC curves}

\begin{figure}%
	\centering
	\includegraphics[trim = 20mm 4mm 12mm 3mm,width=5.6 cm]{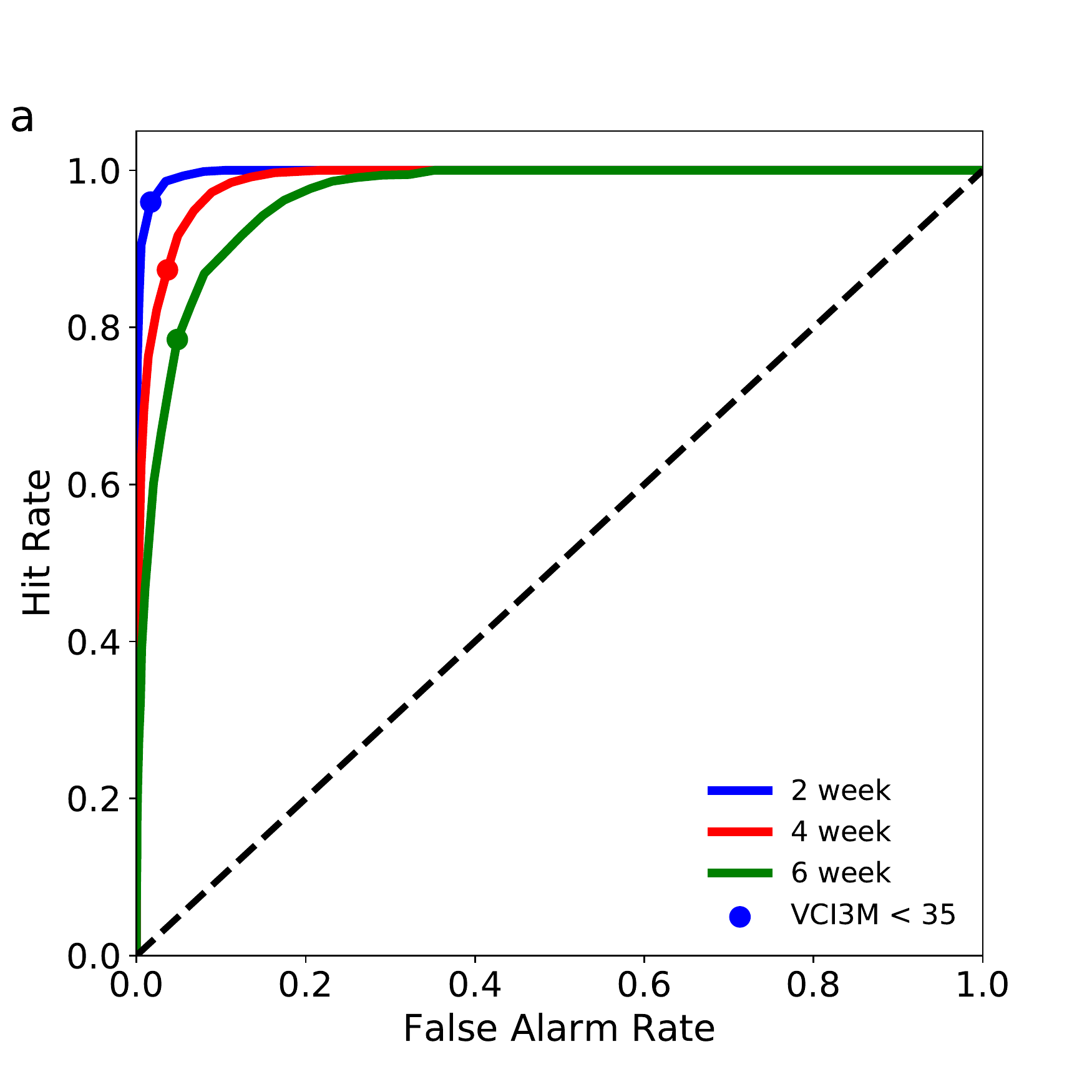}
	\qquad
	\includegraphics[trim = 12mm 4mm 20mm 3mm,width=5.6 cm]{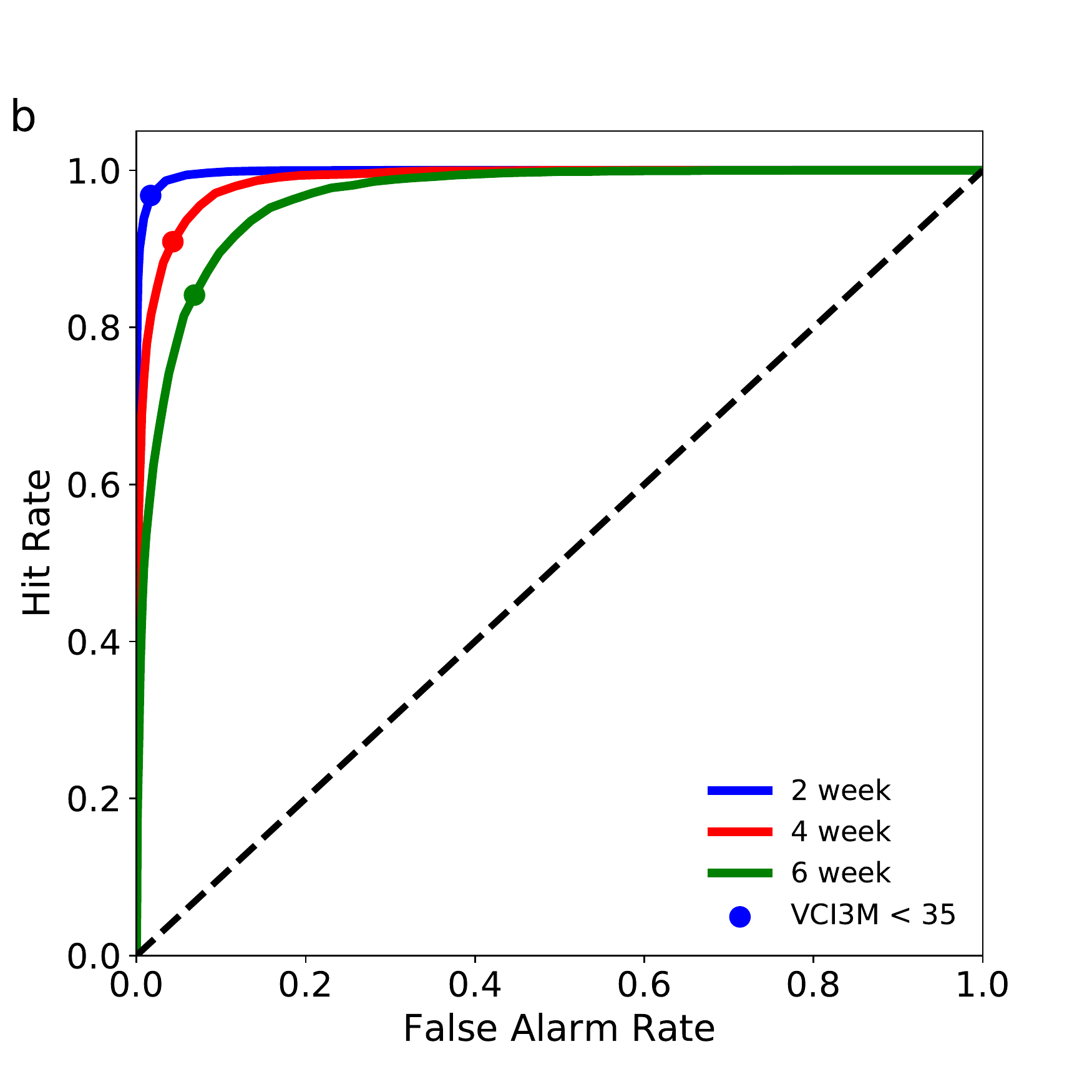}
	\vspace{0.3cm}
	\includegraphics[trim = 20mm 4mm 12mm 3mm,width=5.6 cm]{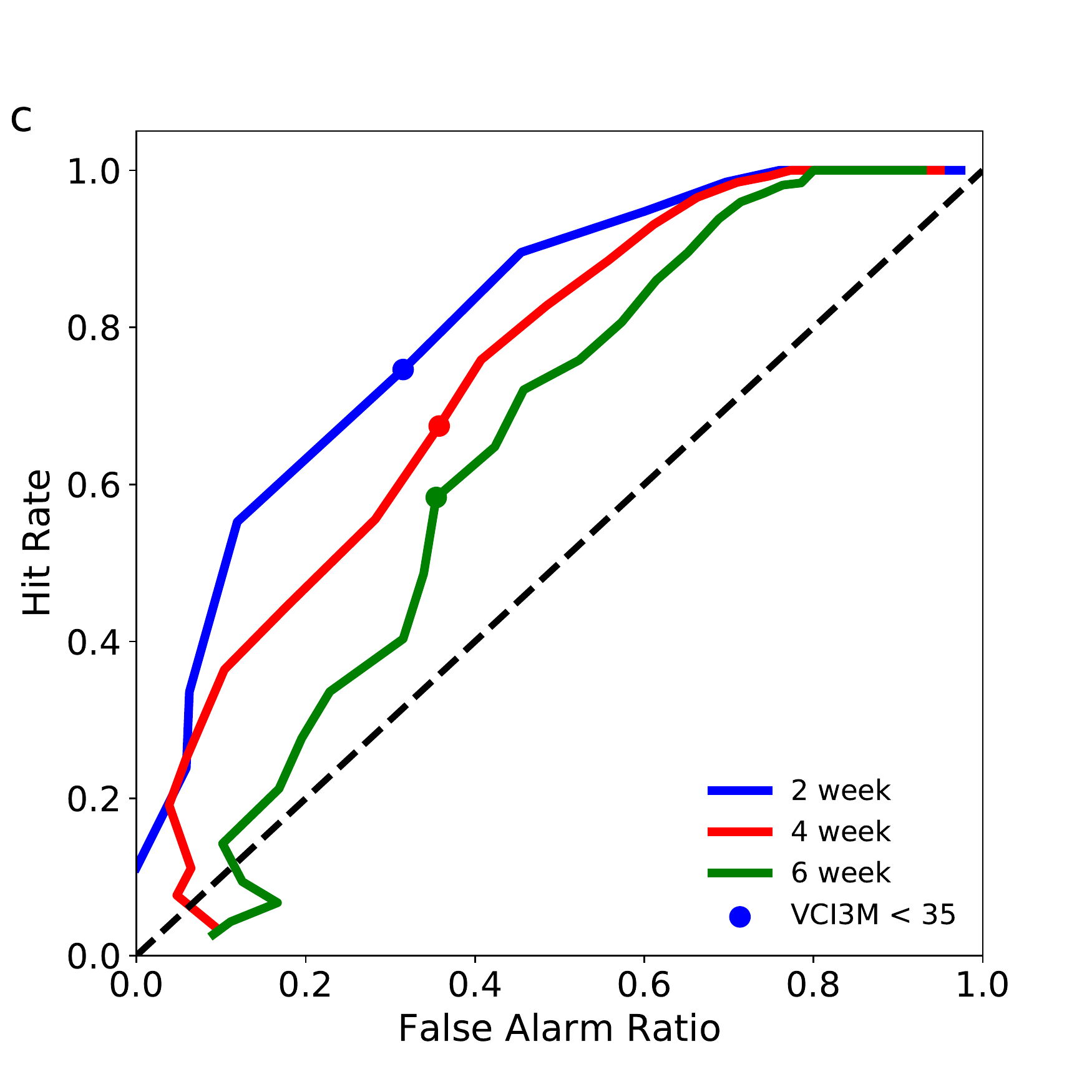}
	\qquad
	\includegraphics[trim = 12mm 4mm 20mm 3mm,width=5.6 cm]{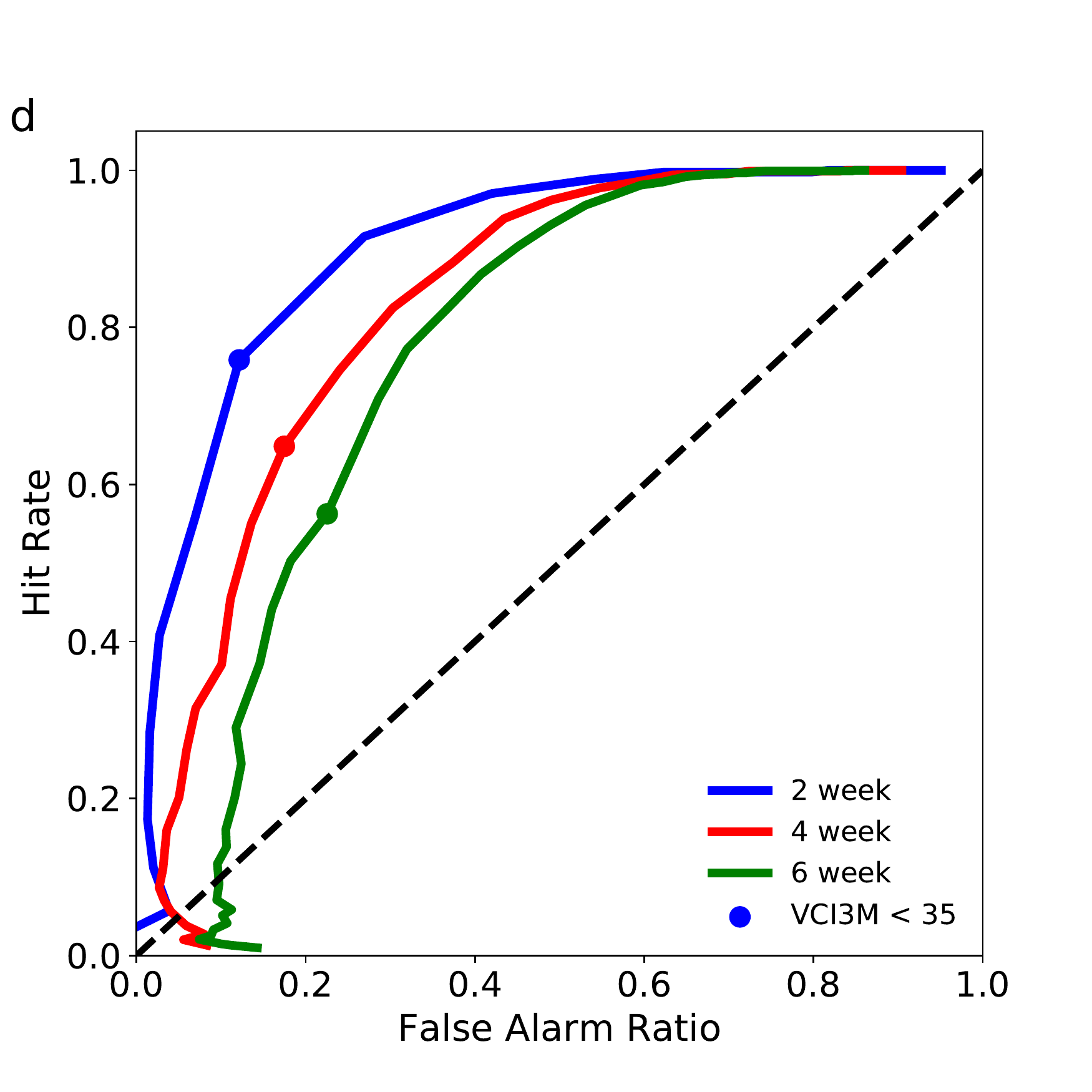}
	\caption{(a) ROC curve for drought detection (VCI3M $<$ 35) for lead times of 2, 4 and 6 weeks using the GP method on Landsat data. (b) ROC curve for drought detection using the AR method on MODIS data. (c, d) Respectively for the GP method on Landsat data and the AR method on the MODIS data, hit rate versus false alarm ratio for forecasting a transition to drought (VCI3M$<35$) given that the vegetation condition is normal (VCI3M$>35$) on the date of the forecast. The curves are plotted from applying different thresholds to convert the continuous forecast into a binary forecast of drought or no drought, see text for details. The shaded circles show the point obtained from forecasting drought when the predicted VCI3M$<$35.}
	\label{fig:ROC_abb}
\end{figure}

To assess the usefulness of the AR and GP methods for drought forecasting, we tested their ability to detect specific drought events, as defined by the NDMA's alert threshold VCI3M$<$35 \cite{rs8040267}. ROC curves were plotted for the detection of VCI3M$<$35 at lead times of two, four and six weeks, see Fig.~\ref{fig:ROC_abb}(a, b). These curves show the probability of predicting a state of drought (VCI3M$<$35) when there will be a state of drought, i.e.~hit rate, against the probability of predicting drought when there will not be drought, i.e.~false alarm rate, for varying binarisation thresholds on the forecast. We further tested the ability to detect the onset of drought, see Fig.~\ref{fig:ROC_abb}(c, d). For this,
the hit rate was defined as the probability of predicting a transition from the normal condition (VCI3M$>$35) to the drought condition (VCI3M$<$35), given that this transition occurs; and the false alarm ratio was defined as the probability that a prediction of this transition is incorrect, given that this transition has been predicted. These curves give an indication that both methods have high skill at forecasting droughts, as measured by the VCI3M, as far as six weeks ahead. The AR forecast appears to do better than the GP forecast at predicting transitions from normal conditions to drought, which can be explained by the tendency of the AR forecast to predict a continuation of the recent trend, while the GP forecast is more likely to predict a reversion to the long term mean value.

The ROC curve performance is not highly dependent on the region (see Table \ref{tab:ROC2}). Even for the wetter Eastern regions, for which observations are sparser due to cloud cover, the hit and false alarm rates only differ by 1 to 2 percentage points compared with those computed across all regions. Further, ROC curves for predicting the NDMA drought categories of severe (10$<$VCI3M$<$20) or extreme (VCI$<$10) drought look similar to those for detecting VCI3M$<$35, see Fig.~\ref{fig:ROCotherdrought} in the Supplementary Material.

\begin{table}
	\small
	\caption{False alarm rate and hit rate (respectively, expressed in percent) for different regions in Kenya and at different lead times. This is based on forecasting drought if the predicted VCI3M is less than 35 (different performances could be obtained with different warning thresholds (see Figure \protect\ref{fig:ROC_abb}. Regions are composed of the following zones: North -- Z1,3 and 5; East --  Z7, 9, 10 and 11 and South -- (Z15 and 18))} \label{tab:ROC2}
	\centering
	\begin{tabular}{l|ccc|ccc} 
		\toprule
		\textbf{Regions} & & \textbf{Landsat GP} & & &\textbf{MODIS AR} \\
		& \textbf{2} & \textbf{4} & \textbf{6} & \textbf{2} & \textbf{4} & \textbf{6} \\
			& \multicolumn{3}{c|}{\textbf{weeks }}& \multicolumn{3}{c}{\textbf{weeks }}\\

		\midrule
		All & 2 \; 96 & 4 \; 87 & 5 \; 78 & 2 \; 97 & 4 \; 91 & \; 7 \; 84\\
		Z24 & 2 \; 99 & 4 \; 91 & 5  \; 82 & 2 \; 98 & 5 \; 94 & \; 8  \; 88\\
		North  & 1 \; 97 & 2 \; 88 & 3  \; 76 & 2 \; 98 & 6 \; 93 & 11 \; 87\\
		East  & 3 \; 94 & 5 \; 85 & 6 \; 77 & 3 \; 97 & 6 \; 91 & 10 \; 85\\
		South   & 1 \; 96 & 3 \; 88 & 4 \; 77 & 2 \; 98 & 6 \; 94 & 11 \; 90\\
		\bottomrule
	\end{tabular}
\end{table}

\section{Discussion} \label{sec:dis}
Droughts are complex and hence inherently difficult to define and measure \cite{MISHRA2010202}. A large number of satellite-based indicators have been developed to identify meteorological, hydrological, and agricultural droughts \cite{zargar2011review,aghakouchak2015remote} with each performing well in space and time to a certain degree \cite{ZHANG201796}. This paper uses two machine-learning methods to provide short-term forecasts of the 3-month VCI (VCI3M), which is used by Kenya's National Drought Management Authority (NDMA) in their drought Early Warning System (EWS). We have investigated the skill and robustness of our forecasts in a number of ways.
Both of our methods showed high sensitivity and specificity for prediction of drought conditions (VCI3M$<$35) as well as the onset of drought, at lead times of 2, 4 and 6 weeks (see Fig.~\ref{fig:ROC_abb}). They also perform better than a persistence forecast (a factor of two in RMSE for VCI3M, and $R^2$ improvement of 0.12). Compared to a similar study that used a Artificial Neural Network model to predict future VCI for four Kenyan counties \cite{Adede2019}, our forecasts provide higher skill, as they showed $R^2$-scores of 0.78 for a 1-month VCI3M forecast compared to 0.95 and 0.94 for our methods.
Moreover, our two methods provide robust results with either dataset (i.e., MODIS and Landsat), and are not impacted by the preprocessing steps. 
Finally, the methods present a high skill in forecasting drought irrespective of the region, the drought category, and the season.

A very important strength of our methods is the high level of skill. It is instructive to understand the origin of this skill; particularly as it may be surprising since the methods are rather simple and do not include other variables (e.g., rainfall, precipitation). \label{review:correlation}
Part of the explanation is that, by using the indicators themselves to determine the forecast, we do track all the factors that impact the vegetation (e.g., disease, soil memory and land-use change) and not just meteorological factors.
Additionally, the natural growth cycles of vegetation and their response to environmental factors introduce temporal correlations (persistence) in the indices which can be exploited in short-range forecasting. 
Furthermore, the VCI3M metric used in this study, and by the NDMA, is additionally smoothed over three months. This smoothing adds temporal dependency, which in turn increases the measured skill. It is important to recognise that this last improvement in skill comes from the inclusion of current data, so it is not a pure forecast and  skill metrics should not be directly compared with skills of e.g. VCI forecasts.  Nevertheless, the high apparent skill is extremely valuable to disaster risk managers who need to make decisions based on uncertain information \cite{wilkinson2018forecasting}.

There are an increasing number of forecasting studies and methodologies for pastures that focus on several indices, with different lead times, and with varying skill \cite{matere2019predictive,Adede2019,Papagiannopoulou,meroni2014early}. Often, new forecast information developed by scientists to help the development and humanitarian sectors enhance disaster preparedness and response goes unused due to a ``usability gap'' between knowledge producers and users \cite{lemos2012narrowing}. In our study, we aim to bridge this gap by focusing on VCI3M, a drought indicator that is currently used by the NDMA to classify drought severity in the arid and semi-arid regions of Kenya. Additionally, decision makers need reliable forecasts to develop robust anticipatory actions in order to mitigate the impacts of drought with limited financial resources \cite{wilkinson2018forecasting}. Our methods provide skillful VCI3M forecasts with detailed information of hit rates and false-alarm ratios, which are often used to define anticipatory actions (see the Red Cross/Red Crescent Forecast-based Financing (FbF) manual for more information\footnote{http://fbf.drk.de/fileadmin/Content/Manual\_FbF/01\_Manual/01\_Manual\_For\_Forecast-Based-Financing.pdf}). Finally, the methodology is also rather simple and easy to implement as it only relies on one data input derived from satellites, which are available globally. The methods can thus be applied everywhere, providing there is sufficient capacity and calibration data. Such co-production strategies allow us to bridge the usability gap \cite{dilling2011creating,lemos2018naturesus} and provides confidence that our forecasting methods may be used.

We have also concentrated on methods that produce accurate short-term forecasts, rather than less-certain, but longer-range forecasts. We can speculate that while the latter might have greater value, the former might be more readily adopted in the monthly county bulletins released by the NDMA. Indeed, the forecasts developed here could, for example, help establish a new drought phase classification (`Early Alert') which, along with adequate preparedness actions developed by the disaster risk managers, would minimise the risk of a worsening drought condition. Anticipatory drought management strategies based on this `Early Alert' could for example focus on livestock vaccination programmes, livestock movement monitoring, or the repair of strategic water sources which enhance the resilience of these communities before a drought hits.

Forecasts alone do not necessarily lead to good anticipatory actions. Whilst acting ahead of disasters is on average more financially effective than responding to an event \cite{Venton2012}, traditionally the humanitarian agencies tend to respond to disasters as financial resources are only available during or after an event. Additionally, due to the uncertainty in the models, anticipatory actions based on such forecasts do raise the risk of ``acting in vain", which may have substantial negative impact on the humanitarian sector in the short term \cite{Lopez2018}. These agencies thus need access to adequate financial resources, e.g. FbF \cite{CoughlandePerez2015} to fund anticipatory actions based on skill-assessed forecast in order to factor in the possible negative consequences of acting in vain. For the forecast methods developed in this study, the chance of acting in vain will be low due to the high level of skill, which will ultimately lower the barriers to uptake.

\section{Caveats and Future Work}
 \label{sec:futurework}

As discussed above, our methods are already sufficiently skillful that they are usable as they stand. However, we have identified some minor limitations and relevant improvements to enhance the functionality, skill, lead-time and impact of our forecasts.

Our analysis has been based on relatively small samples of the available pixels, aggregated, spatially, at the level of the pastoral livelihood zone and county intersections. This limits the localisation specificity of our predictions. Additionally, our methodology using Landsat merges data from various land covers which may reduce accuracy. The processing of all pixels can be achieved within reasonable computational constraints and will allow us to aggregate over specific regions of interest. For example, one could perform the analysis for specific land covers within a county, or for individual grazing units, which would provide greater accuracy and additional functionality.

Our forecasts are currently unavailable, or are less accurate, in periods during or following cloud-cover gaps. More subtly, our validation will have favoured dry season observations, which are less affected by cloud cover, and this will have an impact on the validation of the forecast performance. However, as we found little variation in performance throughout the seasons we do not think these are significant problems. The impact of cloud cover will be reduced when all pixels are processed and aggregated.

As with any machine-learning method, the forecast and its estimated skill are only appropriate for the types of vegetation and environments for which it has been calibrated. The quality that we have obtained in different regions of Kenya gives us some confidence that the skills will not be substantially different.  Nevertheless, the calibration, validation and skill assessment of the forecast will be an essential element of a practical and general tool. Future work should also explore how long a temporal baseline is required for good calibration.  This will, in principle, allow a truly global forecasting tool.

The error estimates we currently provide in our forecasts are derived from the global validation of our performance. They should thus be correct on average but the errors will be overestimated in some situations while, correspondingly, underestimated in others. Future development can provide error estimates that are tailored for the specific conditions and data availability.

The indicators we have chosen are well motivated through their use in the existing EWS operated by NDMA. Identifying the most appropriate and useful indicators of such hazards is the subject of much debate and investigation (see e.g., \cite{zargar2011review,aghakouchak2015remote,ZHANG201796}). But they may not be the most suitable to quantify the relevant socio-economic impacts of droughts \cite{Sutanto2019}. Subject to data availability, similar machine-learning approaches could be applied to more direct socio-economic indicators tracking food insecurity, such as malnutrition, food prices, or livestock condition.

Perhaps the most-significant limitation of our methods is that they are only appropriate for relatively short lead-times. Although a 4-week lead time can be useful, most contingency plans and drought preparedness policies are developed over seasonal timescales. It is thus key to extend this lead-time. While current observations of precipitation and temperature had little impact, including other observed climate variables (e.g., ENSO, sea surface temperature, \cite{Funk2008}) or seasonal climate forecasts may enhance skill and lead times.

Future research will also be required on the effectiveness of the practical implementation of forecasts in EWS \cite{wilkinson2018forecasting}. Clearly-defined triggers (e.g., threshold values based on forecasts, which may vary in time and space) will need to be defined and assessed and optimised against suitable performance metrics. Similarly, effective anticipatory actions need to be defined by the decision makers in relation to these triggers. Adequate policy and institutional arrangements will be needed to allow the various actors to engage and interact with a long-term perspective on risk management. This in turn, requires financial systems that can be accessed based on such forecasts to be able to act across various timescales before the disaster occurs (i.e. Forecast-based-Finance).

\section{Conclusion}

In conclusion we have developed two new forecasting methods which exploit the inherent temporal correlation in vegetation indices to provide highly skillful, short-range forecasts of VCI. The choice of input data, output indicators, simplicity of implementation, and demonstrated skill argues that these methods will be useful for drought early warning systems. We have identified ways this can be improved, but there is clear evidence here that our statistical persistence model provides strong skill over useful lead times. This can be an important contribution to anticipatory drought risk management in Kenya 

\section*{Authors responsibilities}
A.B.B., S.D. and E.S. are lead authors as they contributed equally to the paper and the order of the three names is alphabetical. A.B.B was responsible for developing and running the AR method. S.D. was responsible for developing and running the GP method, and for the accumulation and processing of the Landsat data. E.S. was responsible for the MODIS data accumulation and preprocessing. JMw, SO and PR developed the initial idea and provided feedback throughout. All authors wrote, reviewed and edited the final manuscript. 

\section*{Acknowledgements}
This research was funded by the STFC through the following projects: ``AstroCast: Applying Astronomy Data Analysis to enhance disaster forecasting'' -- grant number ST/R004811/1; and ``STFC Official Development Assistance (ODA) Institutional Award" attached to the same grant; ``A UK-Africa Data Science Network: Capturing the SKA-Driven Data Transformation'' grant number ST/R001898/1; and by the Science for Humanitarian Emergencies and Resilience (SHEAR) consortium project ``Towards Forecast-based Preparedness Action'' (ForPAc, www.forpac.org), grant numbers NE/P000673/1. This project was initiated through pump-priming funding from the University of Sussex's ``Sussex Research'' thematic programme and carried out as part of the interdisciplinary Data Intensive Science Centre at the University of Sussex (DISCUS). We acknowledge early contributions to that pilot work from Peter Hurley, Philip Rooney, Martin Jung, and J\"{o}rn Scharlemann. Martin Todd and Alexander Antonarakis provided useful feedback which improved the manuscript.

\section*{}

\bibliography{astrocast_paper_v4}

\begin{thebibliography}{10}
\expandafter\ifx\csname url\endcsname\relax
  \def\url#1{\texttt{#1}}\fi
\expandafter\ifx\csname urlprefix\endcsname\relax\def\urlprefix{URL }\fi
\expandafter\ifx\csname href\endcsname\relax
  \def\href#1#2{#2} \def\path#1{#1}\fi

\bibitem{Lesk2016}
C.~Lesk, P.~Rowhani, N.~Ramankutty, {Influence of extreme weather disasters on
  global crop production}, Nature 529~(7584) (2016) 84--87.
\newblock \href {https://doi.org/10.1038/nature16467}
  {\path{doi:10.1038/nature16467}}.

\bibitem{Nelson3274}
G.~C. Nelson, H.~Valin, R.~D. Sands, P.~Havl{\'\i}k, H.~Ahammad, D.~Deryng,
  J.~Elliott, S.~Fujimori, T.~Hasegawa, E.~Heyhoe, P.~Kyle, M.~Von~Lampe,
  H.~Lotze-Campen, D.~Mason~d{\textquoteright}Croz, H.~van Meijl, D.~van~der
  Mensbrugghe, C.~M{\"u}ller, A.~Popp, R.~Robertson, S.~Robinson, E.~Schmid,
  C.~Schmitz, A.~Tabeau, D.~Willenbockel,
  \href{https://www.pnas.org/content/111/9/3274}{Climate change effects on
  agriculture: Economic responses to biophysical shocks}, Proceedings of the
  National Academy of Sciences 111~(9) (2014) 3274--3279.
\newblock \href
  {http://arxiv.org/abs/https://www.pnas.org/content/111/9/3274.full.pdf}
  {\path{arXiv:https://www.pnas.org/content/111/9/3274.full.pdf}}, \href
  {https://doi.org/10.1073/pnas.1222465110}
  {\path{doi:10.1073/pnas.1222465110}}.
\newline\urlprefix\url{https://www.pnas.org/content/111/9/3274}

\bibitem{BROWN201531}
M.~E. Brown, V.~Kshirsagar,
  \href{http://www.sciencedirect.com/science/article/pii/S0959378015300248}{Weather
  and international price shocks on food prices in the developing world},
  Global Environmental Change 35 (2015) 31--40.
\newblock \href
  {https://doi.org/https://doi.org/10.1016/j.gloenvcha.2015.08.003}
  {\path{doi:https://doi.org/10.1016/j.gloenvcha.2015.08.003}}.
\newline\urlprefix\url{http://www.sciencedirect.com/science/article/pii/S0959378015300248}

\bibitem{10.1093/rsq/hdr006}
E.~Piguet, A.~Pécoud, P.~de~Guchteneire,
  \href{https://doi.org/10.1093/rsq/hdr006}{Migration and climate change: An
  overview}, Refugee Survey Quarterly 30~(3) (2011) 1--23.
\newblock \href
  {http://arxiv.org/abs/http://oup.prod.sis.lan/rsq/article-pdf/30/3/1/4460951/hdr006.pdf}
  {\path{arXiv:http://oup.prod.sis.lan/rsq/article-pdf/30/3/1/4460951/hdr006.pdf}},
  \href {https://doi.org/10.1093/rsq/hdr006} {\path{doi:10.1093/rsq/hdr006}}.
\newline\urlprefix\url{https://doi.org/10.1093/rsq/hdr006}

\bibitem{Stanke}
C.~Stanke, M.~Kerac, C.~Prudhomme, J.~Medlock, V.~Murray,
  \href{https://www.ncbi.nlm.nih.gov/pubmed/23787891}{Health effects of
  drought: a systematic review of the evidence}, PLoS currents 5 (2013).
\newline\urlprefix\url{https://www.ncbi.nlm.nih.gov/pubmed/23787891}

\bibitem{emdat}
EM-DAT, \href{www.emdat.be}{The International Disaster Database}, Universit\'e
  catholique de Louvain, Belgium, 2019.
\newline\urlprefix\url{www.emdat.be}

\bibitem{Nyong2007}
A.~Nyong, F.~Adesina, B.~Osman~Elasha,
  \href{"https://doi.org/10.1007/s11027-007-9099-0"}{The value of indigenous
  knowledge in climate change mitigation and adaptation strategies in the
  {A}frican sahel}, Mitigation and Adaptation Strategies for Global Change
  12~(5) (2007) 787--797.
\newblock \href {https://doi.org/"10.1007/s11027-007-9099-0"}
  {\path{doi:"10.1007/s11027-007-9099-0"}}.
\newline\urlprefix\url{"https://doi.org/10.1007/s11027-007-9099-0"}

\bibitem{orindi2007}
V.~A. Orindi, A.~Nyong, M.~Herrero, {Pastoral Livelihood Adaptation to Drought
  and Institutional Interventions in Kenya}, Tech. rep. (2007).

\bibitem{Galvin2001}
K.~Galvin, R.~Boone, N.~Smith, S.~Lynn,
  \href{http://www.int-res.com/abstracts/cr/v19/n2/p161-172/}{{Impacts of
  climate variability on East African pastoralists: linking social science and
  remote sensing}}, Climate Research 19 (2001) 161--172.
\newblock \href {https://doi.org/10.3354/cr019161}
  {\path{doi:10.3354/cr019161}}.
\newline\urlprefix\url{http://www.int-res.com/abstracts/cr/v19/n2/p161-172/}

\bibitem{kogan}
F.~Kogan, T.~Adamenko, W.~Guo,
  \href{https://doi.org/10.1080/2150704X.2012.736033}{Global and regional
  drought dynamics in the climate warming era}, Remote Sensing Letters 4~(4)
  (2013) 364--372.
\newblock \href
  {http://arxiv.org/abs/https://doi.org/10.1080/2150704X.2012.736033}
  {\path{arXiv:https://doi.org/10.1080/2150704X.2012.736033}}, \href
  {https://doi.org/10.1080/2150704X.2012.736033}
  {\path{doi:10.1080/2150704X.2012.736033}}.
\newline\urlprefix\url{https://doi.org/10.1080/2150704X.2012.736033}

\bibitem{CoughlandePerez2015}
E.~{Coughlan de Perez}, B.~van~den Hurk, M.~K. van Aalst, B.~Jongman, T.~Klose,
  P.~Suarez,
  \href{https://www.nat-hazards-earth-syst-sci.net/15/895/2015/}{{Forecast-based
  financing: an approach for catalyzing humanitarian action based on extreme
  weather and climate forecasts}}, Natural Hazards and Earth System Sciences
  15~(4) (2015) 895--904.
\newblock \href {https://doi.org/10.5194/nhess-15-895-2015}
  {\path{doi:10.5194/nhess-15-895-2015}}.
\newline\urlprefix\url{https://www.nat-hazards-earth-syst-sci.net/15/895/2015/}

\bibitem{Lopez2018}
A.~Lopez, E.~{Coughlan de Perez}, J.~Bazo, P.~Suarez, B.~van~den Hurk, M.~van
  Aalst, {Bridging forecast verification and humanitarian decisions: A
  valuation approach for setting up action-oriented early warnings}, Weather
  and Climate Extremes (2018).
\newblock \href {https://doi.org/10.1016/j.wace.2018.03.006}
  {\path{doi:10.1016/j.wace.2018.03.006}}.

\bibitem{wilkinson2018forecasting}
E.~Wilkinson, L.~Weing{\"a}rtner, R.~Choularton, M.~Bailey, M.~Todd,
  D.~Kniveton, C.~Cabot~Venton, Forecasting hazards, averting disasters:
  Implementing forecast-based early action at scale, Tech. rep., Overseas
  Development Institute (ODI) (2018).

\bibitem{Venton2012}
C.~Venton, C.~Fitzgibbon, T.~Shitarek, L.~Coulter, O.~Dooley, The economics of
  early response and disaster resilience: lessons from {K}enya and {E}thiopia,
  Independent report (2012).

\bibitem{TozierdelaPoterie2015}
A.~Tozier de~la Poterie, M.-A. Baudoin,
  \href{https://doi.org/10.1007/s13753-015-0053-6}{From {Y}okohama to {S}endai:
  Approaches to participation in international disaster risk reduction
  frameworks}, International Journal of Disaster Risk Science 6~(2) (2015)
  128--139.
\newblock \href {https://doi.org/10.1007/s13753-015-0053-6}
  {\path{doi:10.1007/s13753-015-0053-6}}.
\newline\urlprefix\url{https://doi.org/10.1007/s13753-015-0053-6}

\bibitem{BHUIYAN2006289}
C.~Bhuiyan, R.~Singh, F.~Kogan,
  \href{http://www.sciencedirect.com/science/article/pii/S030324340600016X}{Monitoring
  drought dynamics in the {A}ravalli region ({I}ndia) using different indices
  based on ground and remote sensing data}, International Journal of Applied
  Earth Observation and Geoinformation 8~(4) (2006) 289 -- 302.
\newline\urlprefix\url{http://www.sciencedirect.com/science/article/pii/S030324340600016X}

\bibitem{wmo2015wmo}
WMO, {WMO} guidelines on multi-hazard impact-based forecast and warning
  services (2015).

\bibitem{nhess-2018-26}
F.~Sai, L.~Cumiskey, A.~Weerts, B.~Bhattacharya, R.~Haque~Khan,
  \href{https://www.nat-hazards-earth-syst-sci-discuss.net/nhess-2018-26/}{Towards
  impact-based flood forecasting and warning in {B}angladesh: a case study at
  the local level in {S}irajganj district}, Natural Hazards and Earth System
  Sciences Discussions 2018 (2018) 1--20.
\newblock \href {https://doi.org/10.5194/nhess-2018-26}
  {\path{doi:10.5194/nhess-2018-26}}.
\newline\urlprefix\url{https://www.nat-hazards-earth-syst-sci-discuss.net/nhess-2018-26/}

\bibitem{Sutanto2019}
S.~J. Sutanto, M.~van~der Weert, N.~Wanders, V.~Blauhut, H.~A.~J. {Van Lanen},
  \href{https://doi.org/10.1038/s41467-019-12840-z}{{Moving from drought hazard
  to impact forecasts}}, Nature Communications 10~(1) (2019) 4945.
\newblock \href {https://doi.org/10.1038/s41467-019-12840-z}
  {\path{doi:10.1038/s41467-019-12840-z}}.
\newline\urlprefix\url{https://doi.org/10.1038/s41467-019-12840-z}

\bibitem{KOGAN199591}
F.~Kogan,
  \href{http://www.sciencedirect.com/science/article/pii/027311779500079T}{Application
  of vegetation index and brightness temperature for drought detection},
  Advances in Space Research 15~(11) (1995) 91--100, natural Hazards:
  Monitoring and Assessment Using Remote Sensing Technique".
\newline\urlprefix\url{http://www.sciencedirect.com/science/article/pii/027311779500079T}

\bibitem{rs8040267}
A.~Klisch, C.~Atzberger,
  \href{http://www.mdpi.com/2072-4292/8/4/267}{Operational drought monitoring
  in {K}enya using {MODIS} {{NDVI}} time series}, Remote Sensing 8~(4) (2016).
\newblock \href {https://doi.org/10.3390/rs8040267}
  {\path{doi:10.3390/rs8040267}}.
\newline\urlprefix\url{http://www.mdpi.com/2072-4292/8/4/267}

\bibitem{RULINDA201132}
C.~M. Rulinda, W.~Bijker, A.~Stein,
  \href{http://www.sciencedirect.com/science/article/pii/S1878029611000089}{The
  chlorophyll variability in {M}eteosat derived {NDVI} in a context of drought
  monitoring}, Procedia Environmental Sciences" 3 (2011) 32 -- 37, 1st
  Conference on Spatial Statistics 2011 – Mapping Global Change.
\newblock \href {https://doi.org/https://doi.org/10.1016/j.proenv.2011.02.007}
  {\path{doi:https://doi.org/10.1016/j.proenv.2011.02.007}}.
\newline\urlprefix\url{http://www.sciencedirect.com/science/article/pii/S1878029611000089}

\bibitem{ROJAS2011343}
O.~Rojas, A.~Vrieling, F.~Rembold,
  \href{http://www.sciencedirect.com/science/article/pii/S0034425710002798}{Assessing
  drought probability for agricultural areas in {A}frica with coarse resolution
  remote sensing imagery}, Remote Sensing of Environment 115~(2) (2011) 343 --
  352.
\newblock \href {https://doi.org/https://doi.org/10.1016/j.rse.2010.09.006}
  {\path{doi:https://doi.org/10.1016/j.rse.2010.09.006}}.
\newline\urlprefix\url{http://www.sciencedirect.com/science/article/pii/S0034425710002798}

\bibitem{zargar2011review}
A.~Zargar, R.~Sadiq, B.~Naser, F.~I. Khan, A review of drought indices,
  Environmental Reviews 19~(NA) (2011) 333--349.

\bibitem{quiring2010evaluating}
S.~M. Quiring, S.~Ganesh, Evaluating the utility of the vegetation condition
  index ({VCI}) for monitoring meteorological drought in {T}exas, Agricultural
  and Forest Meteorology 150~(3) (2010) 330--339.

\bibitem{rs8030224}
W.~Jiao, L.~Zhang, Q.~Chang, D.~Fu, Y.~Cen, Q.~Tong,
  \href{https://www.mdpi.com/2072-4292/8/3/224}{Evaluating an enhanced
  vegetation condition index ({VCI}) based on {VIUPD} for drought monitoring in
  the continental united states}, Remote Sensing 8~(3) (2016).
\newblock \href {https://doi.org/10.3390/rs8030224}
  {\path{doi:10.3390/rs8030224}}.
\newline\urlprefix\url{https://www.mdpi.com/2072-4292/8/3/224}

\bibitem{Udelhoven}
T.~Udelhoven, M.~Stellmes, G.~del Barrio, J.~Hill,
  \href{https://doi.org/10.1080/01431160802546829}{Assessment of rainfall and
  {NDVI} anomalies in {S}pain (1989 – 1999) using distributed lag models},
  International Journal of Remote Sensing 30~(8) (2009) 1961--1976.
\newblock \href
  {http://arxiv.org/abs/https://doi.org/10.1080/01431160802546829}
  {\path{arXiv:https://doi.org/10.1080/01431160802546829}}, \href
  {https://doi.org/10.1080/01431160802546829}
  {\path{doi:10.1080/01431160802546829}}.
\newline\urlprefix\url{https://doi.org/10.1080/01431160802546829}

\bibitem{meroni2014early}
M.~Meroni, D.~Fasbender, F.~Kayitakire, G.~Pini, F.~Rembold, F.~Urbano,
  M.~Verstraete, Early detection of biomass production deficit hot-spots in
  semi-arid environment using {FAPAR} time series and a probabilistic approach,
  Remote Sensing of Environment 142 (2014) 57--68.

\bibitem{Zambrano:2018}
F.~Zambrano, A.~Vrieling, A.~Nelson, M.~Meroni, T.~Tadesse,
  \href{http://www.sciencedirect.com/science/article/pii/S0034425718304541}{Prediction
  of drought-induced reduction of agricultural productivity in {C}hile from
  {MODIS}, rainfall estimates, and climate oscillation indices}, Remote Sensing
  of Environment 219 (2018) 15--30.
\newblock \href {https://doi.org/https://doi.org/10.1016/j.rse.2018.10.006}
  {\path{doi:https://doi.org/10.1016/j.rse.2018.10.006}}.
\newline\urlprefix\url{http://www.sciencedirect.com/science/article/pii/S0034425718304541}

\bibitem{vrieling2016early}
A.~Vrieling, M.~Meroni, A.~G. Mude, S.~Chantarat, C.~C. Ummenhofer, K.~C.
  de~Bie, Early assessment of seasonal forage availability for mitigating the
  impact of drought on {E}ast {A}frican pastoralists, Remote sensing of
  environment 174 (2016) 44--55.

\bibitem{matere2019predictive}
J.~Matere, P.~Simpkin, J.~Angerer, E.~Olesambu, S.~Ramasamy, F.~Fasina,
  Predictive livestock early warning system ({PLEWS}): Monitoring forage
  condition and implications for animal production in {K}enya, Weather and
  Climate Extremes (2019) 100209.

\bibitem{gpm}
C.~E. Rasmussen, C.~K.~I. Williams, Gaussian processes for machine learning
  (2006).

\bibitem{Hamilton94}
J.~Hamilton, Time series analysis, Princeton University Press, Princeton, NJ,
  1994.

\bibitem{955315}
S.~{Brahim-Belhouari}, J.~M. {Vesin}, Bayesian learning using {G}aussian
  process for time series prediction, in: Proceedings of the 11th IEEE Signal
  Processing Workshop on Statistical Signal Processing (Cat. No.01TH8563),
  2001, pp. 433--436.
\newblock \href {https://doi.org/10.1109/SSP.2001.955315}
  {\path{doi:10.1109/SSP.2001.955315}}.

\bibitem{Chandola2010SCALABLETS}
V.~Chandola, R.~R. Vatsavai,
  \href{https://onlinelibrary.wiley.com/doi/abs/10.1002/sam.10129}{A scalable
  {G}aussian process analysis algorithm for biomass monitoring}, Statistical
  Analysis and Data Mining: The ASA Data Science Journal 4~(4) (2011) 430--445.
\newline\urlprefix\url{https://onlinelibrary.wiley.com/doi/abs/10.1002/sam.10129}

\bibitem{7487896}
G.~Camps-Valls, J.~Verrelst, J.~Munoz-Mari, V.~Laparra, F.~Mateo-Jimenez,
  J.~Gomez-Dans, A survey on {G}aussian processes for earth-observation data
  analysis: A comprehensive investigation, IEEE Geoscience and Remote Sensing
  Magazine 4~(2) (2016) 58--78.
\newblock \href {https://doi.org/10.1109/MGRS.2015.2510084}
  {\path{doi:10.1109/MGRS.2015.2510084}}.

\bibitem{rs11050481}
D.~Upreti, W.~Huang, W.~Kong, S.~Pascucci, S.~Pignatti, X.~Zhou, H.~Ye,
  R.~Casa, \href{http://www.mdpi.com/2072-4292/11/5/481}{A comparison of hybrid
  machine learning algorithms for the retrieval of wheat biophysical variables
  from {S}entinel-2}, Remote Sensing 11~(5) (2019).
\newblock \href {https://doi.org/10.3390/rs11050481}
  {\path{doi:10.3390/rs11050481}}.
\newline\urlprefix\url{http://www.mdpi.com/2072-4292/11/5/481}

\bibitem{Asoka:2015}
A.~Asoka, V.~Mishra, Prediction of vegetation anomalies to improve food
  security and water management in {I}ndia, Geophysical Research Letters
  42~(13) (2015) 5290--5298.
\newblock \href {https://doi.org/10.1002/2015GL063991}
  {\path{doi:10.1002/2015GL063991}}.

\bibitem{Papagiannopoulou}
C.~Papagiannopoulou, D.~Gonzalez~Miralles, S.~Decubber, M.~Demuzere,
  N.~Verhoest, W.~A. Dorigo, W.~Waegeman,
  \href{http://dx.doi.org/10.5194/gmd-10-1945-2017}{A non-linear
  {G}ranger-causality framework to investigate climate-vegetation dynamics},
  GEOSCIENTIFIC MODEL DEVELOPMENT 10~(5) (2017) 1945--1960.
\newline\urlprefix\url{http://dx.doi.org/10.5194/gmd-10-1945-2017}

\bibitem{UNDP2013}
UNDP, \href{http://gefonline.org/projectDetailsSQL.cfm?projID=3792.}{{K}enya
  – adaptation to climate change in arid lands ({KACCAL})-{K}enya case
  study}, Tech. rep. (2013).
\newline\urlprefix\url{http://gefonline.org/projectDetailsSQL.cfm?projID=3792.}

\bibitem{FAO2014}
FAO, \href{www.fao.org/publications}{{Food and Agriculture Organization(FAO)
  Country programming framework for {K}enya (2014-2017)}}, Tech. rep. (2014).
\newline\urlprefix\url{www.fao.org/publications}

\bibitem{Behnke2011}
R.~Behnke, D.~Muthami,
  \href{https://cgspace.cgiar.org/bitstream/handle/10568/24972/IGAD{\_}LPI{\_}WP{\_}03-11.pdf?sequence=1{\&}isAllowed=y}{{The
  Contribution of Livestock to the Kenyan Economy}}, Tech. rep. (2011).
\newline\urlprefix\url{https://cgspace.cgiar.org/bitstream/handle/10568/24972/IGAD{\_}LPI{\_}WP{\_}03-11.pdf?sequence=1{\&}isAllowed=y}

\bibitem{Savitzky1964}
A.~Savitzky, M.~J.~E. Golay,
  \href{http://pubs.acs.org/doi/abs/10.1021/ac60214a047}{{Smoothing and
  Differentiation of Data by Simplified Least Squares Procedures.}}, Analytical
  Chemistry 36~(8) (1964) 1627--1639.
\newblock \href {https://doi.org/10.1021/ac60214a047}
  {\path{doi:10.1021/ac60214a047}}.
\newline\urlprefix\url{http://pubs.acs.org/doi/abs/10.1021/ac60214a047}

\bibitem{MISHRA2010202}
A.~K. Mishra, V.~P. Singh,
  \href{http://www.sciencedirect.com/science/article/pii/S0022169410004257}{A
  review of drought concepts}, Journal of Hydrology 391~(1) (2010) 202--216.
\newblock \href {https://doi.org/https://doi.org/10.1016/j.jhydrol.2010.07.012}
  {\path{doi:https://doi.org/10.1016/j.jhydrol.2010.07.012}}.
\newline\urlprefix\url{http://www.sciencedirect.com/science/article/pii/S0022169410004257}

\bibitem{aghakouchak2015remote}
A.~AghaKouchak, A.~Farahmand, F.~Melton, J.~Teixeira, M.~Anderson, B.~D.
  Wardlow, C.~Hain, Remote sensing of drought: Progress, challenges and
  opportunities, Reviews of Geophysics 53~(2) (2015) 452--480.

\bibitem{ZHANG201796}
L.~Zhang, W.~Jiao, H.~Zhang, C.~Huang, Q.~Tong,
  \href{http://www.sciencedirect.com/science/article/pii/S0034425716304813}{Studying
  drought phenomena in the continental united states in 2011 and 2012 using
  various drought indices}, Remote Sensing of Environment 190 (2017) 96--106.
\newblock \href {https://doi.org/https://doi.org/10.1016/j.rse.2016.12.010}
  {\path{doi:https://doi.org/10.1016/j.rse.2016.12.010}}.
\newline\urlprefix\url{http://www.sciencedirect.com/science/article/pii/S0034425716304813}

\bibitem{Adede2019}
C.~Adede, R.~Oboko, P.~W. Wagacha, C.~Atzberger,
  \href{https://www.mdpi.com/2072-4292/11/9/1099}{{A Mixed Model Approach to
  Vegetation Condition Prediction Using Artificial Neural Networks (ANN): Case
  of Kenya's Operational Drought Monitoring}}, Remote Sensing 11~(9) (2019)
  1099.
\newblock \href {https://doi.org/10.3390/rs11091099}
  {\path{doi:10.3390/rs11091099}}.
\newline\urlprefix\url{https://www.mdpi.com/2072-4292/11/9/1099}

\bibitem{lemos2012narrowing}
M.~C. Lemos, C.~J. Kirchhoff, V.~Ramprasad, Narrowing the climate information
  usability gap, Nature climate change 2~(11) (2012) 789.

\bibitem{dilling2011creating}
L.~Dilling, M.~C. Lemos, Creating usable science: Opportunities and constraints
  for climate knowledge use and their implications for science policy, Global
  environmental change 21~(2) (2011) 680--689.

\bibitem{lemos2018naturesus}
M.~C. Lemos, J.~C. Arnott, N.~M. Ardoin, K.~Baja, A.~T. Bednarek, A.~Dewulf,
  C.~Fieseler, K.~A. Goodrich, K.~Jagannathan, N.~Klenk, et~al., To co-produce
  or not to co-produce, Nature Sustainability 1~(12) (2018) 722.

\bibitem{Funk2008}
C.~Funk, M.~D. Dettinger, J.~C. Michaelsen, J.~P. Verdin, M.~E. Brown,
  M.~Barlow, A.~Hoell,
  \href{http://www.pnas.org/content/105/32/11081.abstract}{{Warming of the
  {I}ndian {O}cean threatens eastern and southern {A}frican food security but
  could be mitigated by agricultural development}}, Proceedings of the National
  Academy of Sciences 105~(32) (2008) 11081 LP -- 11086.
\newblock \href {https://doi.org/10.1073/pnas.0708196105}
  {\path{doi:10.1073/pnas.0708196105}}.
\newline\urlprefix\url{http://www.pnas.org/content/105/32/11081.abstract}

\bibitem{royetal}
D.~Roy, M.~Wulder, T.~Loveland, C.~Woodcock, R.~Allen, M.~Anderson, D.~Helder,
  J.~Irons, D.~Johnson, R.~Kennedy, T.~Scambos, C.~Schaaf, J.~Schott, Y.~Sheng,
  E.~Vermote, A.~Belward, R.~Bindschadler, W.~Wohen, F.~Gao, Z.~Zhu, Landsat-8:
  Science and product vision for terrestrial global change research, Remote
  Sensing of Environment 2014 (2014) 154--172.
\newblock \href {https://doi.org/10.1016/j.rse.2014.02.001}
  {\path{doi:10.1016/j.rse.2014.02.001}}.

\bibitem{Schaaf2015}
C.~Schaaf, Z.~Wang,
  \href{https://lpdaac.usgs.gov/dataset{\_}discovery/modis/modis{\_}products{\_}table/mcd43a4{\_}v006}{{MCD43A4
  MODIS/Terra+Aqua BRDF/Albedo Nadir BRDF Adjusted Ref Daily L3 Global - 500m
  V006}} (2015).
\newblock \href {https://doi.org/10.5067/MODIS/MCD43A4.006}
  {\path{doi:10.5067/MODIS/MCD43A4.006}}.
\newline\urlprefix\url{https://lpdaac.usgs.gov/dataset{\_}discovery/modis/modis{\_}products{\_}table/mcd43a4{\_}v006}

\bibitem{pmlr-v28-duvenaud13}
D.~Duvenaud, J.~Lloyd, R.~Grosse, J.~Tenenbaum, G.~Zoubin,
  \href{http://proceedings.mlr.press/v28/duvenaud13.html}{Structure discovery
  in nonparametric regression through compositional kernel search}, in:
  S.~Dasgupta, D.~McAllester (Eds.), Proceedings of the 30th International
  Conference on Machine Learning, Vol.~28 of Proceedings of Machine Learning
  Research, PMLR, Atlanta, Georgia, USA, 2013, pp. 1166--1174.
\newline\urlprefix\url{http://proceedings.mlr.press/v28/duvenaud13.html}

\bibitem{Hoffman:2013}
M.~D. Hoffman, D.~M. Blei, C.~Wang, J.~Paisley, Stochastic variational
  inference, J. Mach. Learn. Res. 14~(1) (2013) 1303–1347.

\bibitem{bg-10-4055-2013}
S.~Kandasamy, F.~Baret, A.~Verger, P.~Neveux, M.~Weiss,
  \href{https://www.biogeosciences.net/10/4055/2013/}{A comparison of methods
  for smoothing and gap filling time series of remote sensing observations -
  application to {MODIS} {LAI} products}, Biogeosciences 10~(6) (2013)
  4055--4071.
\newblock \href {https://doi.org/10.5194/bg-10-4055-2013}
  {\path{doi:10.5194/bg-10-4055-2013}}.
\newline\urlprefix\url{https://www.biogeosciences.net/10/4055/2013/}

\bibitem{Weiss2014}
D.~J. Weiss, P.~M. Atkinson, S.~Bhatt, B.~Mappin, S.~I. Hay, P.~W. Gething,
  \href{https://www.sciencedirect.com/science/article/pii/S0924271614002512}{{An
  effective approach for gap-filling continental scale remotely sensed
  time-series}}, ISPRS Journal of Photogrammetry and Remote Sensing 98 (2014)
  106--118.
\newblock \href {https://doi.org/10.1016/J.ISPRSJPRS.2014.10.001}
  {\path{doi:10.1016/J.ISPRSJPRS.2014.10.001}}.
\newline\urlprefix\url{https://www.sciencedirect.com/science/article/pii/S0924271614002512}

\bibitem{Cao:2018}
R.~Cao, Y.~Chen, M.~Shen, J.~Chen, J.~Zhou, C.~Wang, W.~Yang,
  \href{http://www.sciencedirect.com/science/article/pii/S0034425718303985}{A
  simple method to improve the quality of {{NDVI}} time-series data by
  integrating spatiotemporal information with the {S}avitzky-{G}olay filter},
  Remote Sensing of Environment 217 (2018) 244 -- 257.
\newblock \href {https://doi.org/https://doi.org/10.1016/j.rse.2018.08.022}
  {\path{doi:https://doi.org/10.1016/j.rse.2018.08.022}}.
\newline\urlprefix\url{http://www.sciencedirect.com/science/article/pii/S0034425718303985}

\end{thebibliography}

\newpage
\listoffigures

\listoftables

\newpage
  
\renewcommand\appendixname{Supplementary Material }
\setcounter{figure}{0}    
\setcounter{table}{0}
\setcounter{page}{1}

\appendix

\section*{Supplementary Material}

\section{Data selection and comparison of datasets} \label{sec:datasets}

\subsection{Landsat}
Landsat-5, 7 and 8 \cite{royetal} red and near infrared (NIR) surface reflectances and quality assessment (QA) data over the 10 pastoral livelihood zones of Kenya, from January 1st, 2000 to February 1st, 2019, were obtained using the United States Geological Survey (USGS) EarthExplorer. Specifically, data from 1\,000 pixels within each region were drawn from the Level-1 Precision Terrain (L1TP) processed dataset, which has well-characterized radiometry and is inter-calibrated across the different Landsat sensors. The spatial resolution of these data is 30m and the repeat interval is 16 days. Using the QA data, observations classified as clear from clouds or cloud shadows were kept. Pixels with fewer than half of the observations over the full time period were discarded (and replaced with an alternative random selection, with a few exceptions, see Fig.~\ref{fig:R2plots}). 
The surface reflectances were combined to obtain 
NDVI.

\subsection{MODIS}

NDVI data were also gathered from the surface reflectances obtained from the daily, 500-meter resolution MODIS Terra/Aqua Nadir BRDF-Adjusted Reflectance product MCD43A4,v006 \cite{Schaaf2015}. Data from February 22nd, 2000 up to February 1st, 2019 were acquired via the NASA Land Processes Distributed Active Archive Center. QA maps files with binary quality flags were used to remove poor quality data resulting from cloud or unreliable BRDF corrections. Data were drawn from 100  pixels within each region, out of those that had been  identified  as grassland by the MODIS land cover classification maps (MCD12Q1,v006).

\subsection{Comparison of the two datasets} 

\begin{table}[!hb]
	\small
	\caption{Table comparing Landsat and MODIS products}\label{tab:sm1}
	\label{tab:comp}
	\centering
	\begin{tabular}{p{2cm}p{4cm}p{4cm}}
		\toprule
		\textbf{Feature}	& \textbf{Landsat} & \textbf{MODIS}\\
		\midrule
		\textbf{Spatial \newline Resolution} & 	High resolution at 30\,m 	&   Medium resolution ranging from 250\,m to 1\,km \\
		\textbf{Temporal \newline Resolution}	&  16-day sampling (8-day when both Landsat-7 and 8 are used	&   Daily sampling  monitoring dynamic variables  \\ 
		\textbf{Quality} &Cloud coverage at 30\,m & Cloud coverage at 500\,m \\
		
		\bottomrule
	\end{tabular}
\end{table}

The key differences between the two datasets are the spatial and temporal resolutions, see Table \ref{tab:comp}. The Landsat data have higher spatial resolution, whilst the MODIS data have higher temporal resolution. Since forecasting was being attempted at the level of large scale regions (livelihood zone and county intersections), and at a weekly temporal resolution, the expectation was that the MODIS data would have advantages,  assuming individual Landsat and MODIS observations have similar signal-to-noise ratios. The processed MODIS time series with weekly observations have less measurement noise because they are composites of 7 daily observations (that themselves are 16-day composites of measurements taken every 1-2 days), whereas the processed Landsat time series are derived from more temporally sparse data (up to 3 different Landsat missions, each yielding one observation every 16 days). Landsat data would have advantages in different applications where forecasts on smaller spatial scales are required. The Landsat data also has the advantage that the quality flags and cloud masks are defined on smaller scales.    

\begin{figure} 
	\centering
	\includegraphics[trim = 0mm 0mm 5mm 20mm,width=12.5 cm]{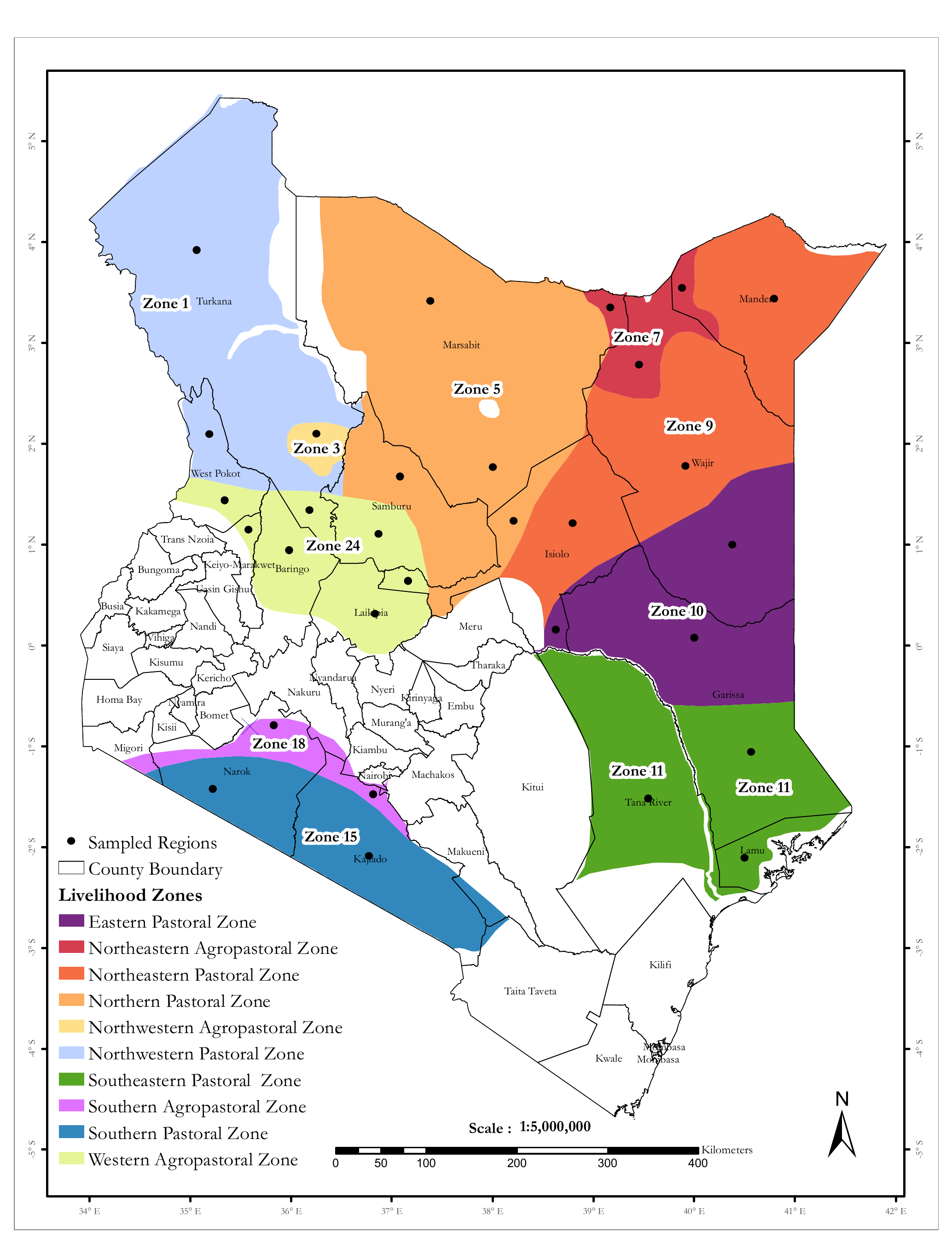}
	\caption{Map of Kenya showing the livelihood zones from which pixels were sampled. }  \label{fig:l_zone}
\end{figure}

The differences between the MODIS and Landsat datasets produced slightly different `true' aggregate time series on which to assess the interpolation and forecasting methods. In addition to the different temporal resolution of the observations supplying the final time series, the MODIS data were aggregated across 100 random grassland pixels from each region, whereas the 1\,000 Landsat pixels analysed were randomly distributed over the whole of each region. In choosing how many pixels to analyse per region, there is a trade-off between using a larger number of pixels for higher accuracy, and a smaller number of pixels for lower computational cost. Fewer MODIS pixels were used than Landsat pixels since they correspond to larger spatial regions. Both these choices of number of pixels should be sufficient for high accuracy of results, since for Landsat data the $R^2$-score comparing the average of all pixels from a region with the average of 100 or 1\,000 random pixels was 0.990 and 0.9993 respectively. The MODIS grassland classification was not available at Landsat resolution, thus unambiguous classification of the smaller Landsat pixels was not possible. This is unlikely to have made much difference to pixel selection, given that the pastoral livelihood zones are mostly grasslands (Fig. \ref{fig:l_zone}). 

\section{Further details on preprocessing}

\subsection{Gaussian process modelling} \label{sec:GPexplain}
A Gaussian Process is a probabilistic model defined as a collection of random variables for which any finite subset has a joint Gaussian distribution \cite{gpm}. Formally, for the present application of interpolation or extrapolation of a time series, with observation at time $t$ denoted by $X_t$, the model is

\begin{align}
    &X_t\sim \mathcal{N}\left[ Y(t),\sigma_r^2\right]\,,\\
    &Y(t) \sim \mathcal{GP}\left[ m(t),k(t,t^{\prime})\right]\,.
\end{align}
Here $Y(t)$ is the true value of the observed index, and the measurement noise is $\sigma_r$, so that an observation $X_t$ is a normal random variable with mean $Y(t)$ and standard deviation $\sigma_r$. The true values $Y(t)$ are also normally distributed, with the mean at time $t$ given by the mean function $m(t)$, and the covariance between values at times $t$ and $t^{\prime}$ given by the kernel function $k(t,t^{\prime})$. To carry out interpolation or extrapolation from a time series, existing data are used to fit the mean, $m$, kernel, $k$, and measurement noise $\sigma_r$, and then expected values are produced for the desired times, based on the obtained fit.

For gap-filling on individual Landsat pixel NDVI time series, the model was determined as follows, using the \texttt{Pyro} programming package for \texttt{Python}. The mean, $m(t)$, was assumed to be constant, and the mean of the whole time series. To determine the kernel, Compositional Kernel Search \cite{pmlr-v28-duvenaud13} was used. Specifically, a search through all the following kernels, and products and sums of pairs of them was carried out: Linear, Radial Basis Function (RBF), Periodic (with period $p$ set to one year), Rational Quadratic, and Matern. The highest marginal likelihood was achieved by Radial Basis Function (RBF) plus Periodic $(k_{\mathrm{RBF}} + k_{\mathrm{P}})$, so this combination was selected as the kernel:

\begin{align}
	k_{\rm{RBF}}(t,t') &= \sigma_{\rm{RBF}}^2 \exp{\Big( -0.5 \frac{|t-t'|^2}{l_{\rm{RBF}}^2} \Big) }\,,\\
	k_{\rm{P}}(t,t')&=\sigma_{\rm{P}}^2 \exp \left( -2 \frac{ \sin^2(\pi|t-t'|/p)}{l_{\mathrm{P}}^2} \right)\,.
\end{align}
There were thus 5 parameters to fit for each time series $(\sigma_r,\sigma_{\rm{RBF}},l_{\rm{RBF}},\sigma_{\rm{P}},l_{\mathrm{P}})$. These were learned using Stochastic Variational Inference \cite{Hoffman:2013}.

For the forecasting on the aggregated NDVI anomaly and VCI3M, a pure Radial Basis Function kernel was used, since for these anomaly indices, the periodic component is not present, see Section \ref{sec:methods} in the main manuscript.

\subsection{Gap-filling for MODIS}\label{sec:Lmax}

Interpolation of gaps in the raw MODIS time series was not carried out when the length of the gap was longer than a certain maximum, $L_{\rm{max}}$. In choosing $L_{\rm{max}}$, a trade off between quality and quantity of remaining observations had to be made: a small $L_{\rm{max}}$ would lead to fewer forecasts being attempted, but interpolations closer to the ground truth, while a large $L_{\rm{max}}$ would lead to more forecasts being attempted, but with these forecasts being assessed against interpolations that are potentially far from the ground truth. The choice $L_{\rm{max}}=6\, \rm{weeks}$ was made, after exploring a range of values and finding $R^2$-score to be not sensitive to the precise choice within the range between 4 and 8 weeks, see Table \ref{tab:IL choices}. 
Note that interpolation on the Landsat data was carried out for all gaps, since the GP interpolation method makes use of the entire time series, and interpolated values within a long interpolation take values close to the seasonal mean.

\begin{table}
	\caption{Comparison of outcomes for different choices of maximum allowed interpolation length $L_{\mathrm{max}}$ on the MODIS data. $R^2$-score of 4 week AR forecast} and the percentage of the time that it was possible to make a forecast, for $L_{\mathrm{max}}=4$, 6, and 8 weeks. Numbers show the median across all regions. \label{tab:IL choices}
	\centering
	\begin{tabular}{ccc} 
		\toprule
		\textbf{$L_{\mathrm{max}}$ (weeks)}   & \textbf{$R^2$-score}    & \textbf{Forecasts attempted (\%)}\\
		\midrule
		4  & 0.60 & 84 \\
		6  & 0.58 & 93 \\
		8 & 0.63 & 98 \\
		\bottomrule
	\end{tabular}
\end{table}

Due to the presence of non-interpolated gaps in the MODIS time series, there were weeks when a forecast assessment was not carried out on these data. The criteria for being able to do AR forecasting on these data were: (i) the three most recent weekly aggregated observations had to be present, since these are required for making a prediction; (ii) there had to be an aggregated observation present for the week being forecast, so the quality of the prediction could be assessed.\footnote{GP forecasting was still possible when (i) failed, but was also not carried out in that case, since performance would have been worse than usual in this case.}

\subsection{Comparison of other possible gap-filling methods} \label{sec:compare}
Various gap-filling methods have been used to deal with missing values resulting from the presence of clouds and atmospheric aerosols. These methods are based on either spatial information, temporal information or some combination of both spatial and temporal information \cite{bg-10-4055-2013, Weiss2014}. Temporal interpolation was chosen given that spatial interpolation methods suffer from the fact that there are frequently clouds over Kenya that cover large groups of neighbouring pixels (although a possible alternative, not considered here, would be to make use of other pixels that historically behave similarly in time \cite{Cao:2018}.

The performance of the temporal gap-filling methods employed, compared with alternative temporal gap-filling methods, was tested by removing observations, applying the method, and then comparing the interpolated observations with the removed observations. GP interpolation and linear, quadratic and cubic polynomial interpolation methods were tested, on both the Landsat and MODIS datasets. $R^2$-scores were obtained for using the interpolated values to predict the `true' values for the missing observations.


For the Landsat data, one randomly chosen observation between 1/1/2014 and 1/2/2019 was removed from each of 2000 randomly selected individual pixel time series.
From the MODIS data, 2000 random individual pixel NDVI time series (1/1/2014 to 1/2/2019) were chosen. 20 randomly selected NDVI values were dropped from each of the time series and the various gap-filling methods were used to interpolate the dropped values. The results for Landsat are shown in Table \ref{tab:comp_int}, and for MODIS in Table \ref{tab:comp_int2}. Note that with these methods, the random samples are more likely to come from periods when there are not many gaps. It is an assumption that the results are valid across all periods.

\begin{table}
	\caption{Comparison of GP method with commonly used interpolation methods as candidates for gap-filling on Landsat data. At the pixel level a random observation was removed, and then interpolated with each of the listed methods.} \label{tab:comp_int}
	\centering
	\begin{tabular}{lc} 
		\toprule
		\textbf{Method}  & \textbf{$R^2$-score} \\
		\midrule
		GP & 0.67 \\
		Linear & 0.53 \\
		Quadratic & -0.07 \\
		Cubic & -1.92 \\
		Last value & 0.34 \\
		Mean value & 0.0 \\
		\bottomrule
	\end{tabular}
\end{table}

\begin{figure} 
	\centering
	\includegraphics[trim = 30mm 5mm 5mm 15mm,width=8cm]{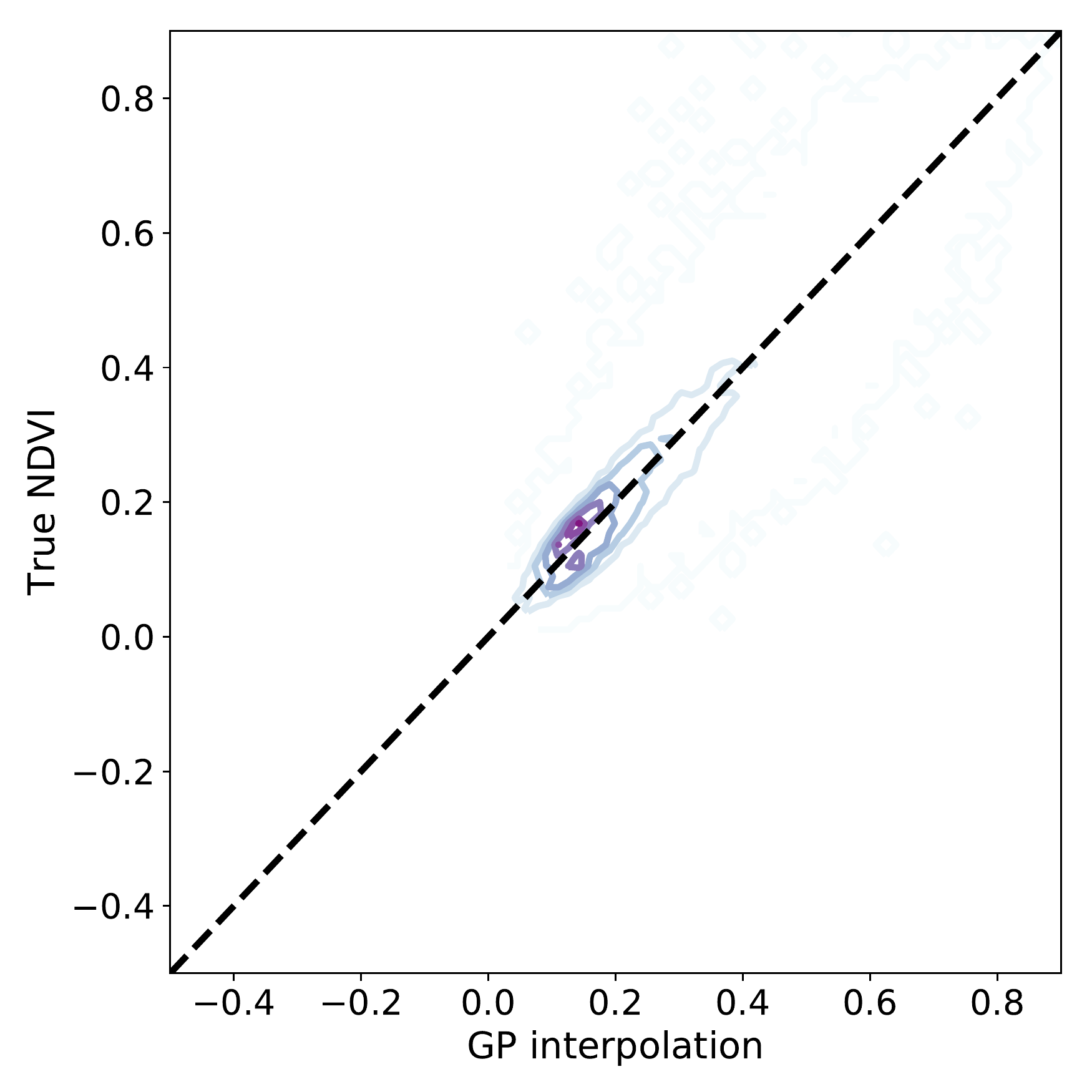}
	\caption{Contour plot of Landsat observed and predicted NDVI values from the GP interpolation.} \label{fig:interp_GP}
\end{figure}

For the Landsat data, the GP method achieved the highest $R^2$-score, thus showing its utility, and justifying our choosing it. The $R^2$-score of 0.67, achieved by the GP method, is close to the $R^2$-score of 0.76 which is obtained from using one Landsat observation to predict a second Landsat observation from the same 16-day observation period (of which there were instances in the data). Fig.~\ref{fig:interp_GP} shows a contour plot of the true versus interpolated NDVI observations using this method. This plot shows that the method doesn't introduce any biases- the slope and intercept are approximately 1 and 0 respectively.

\begin{figure}[H]
	\centering
	\includegraphics[trim = 30mm 5mm 5mm 15mm,width=8cm]{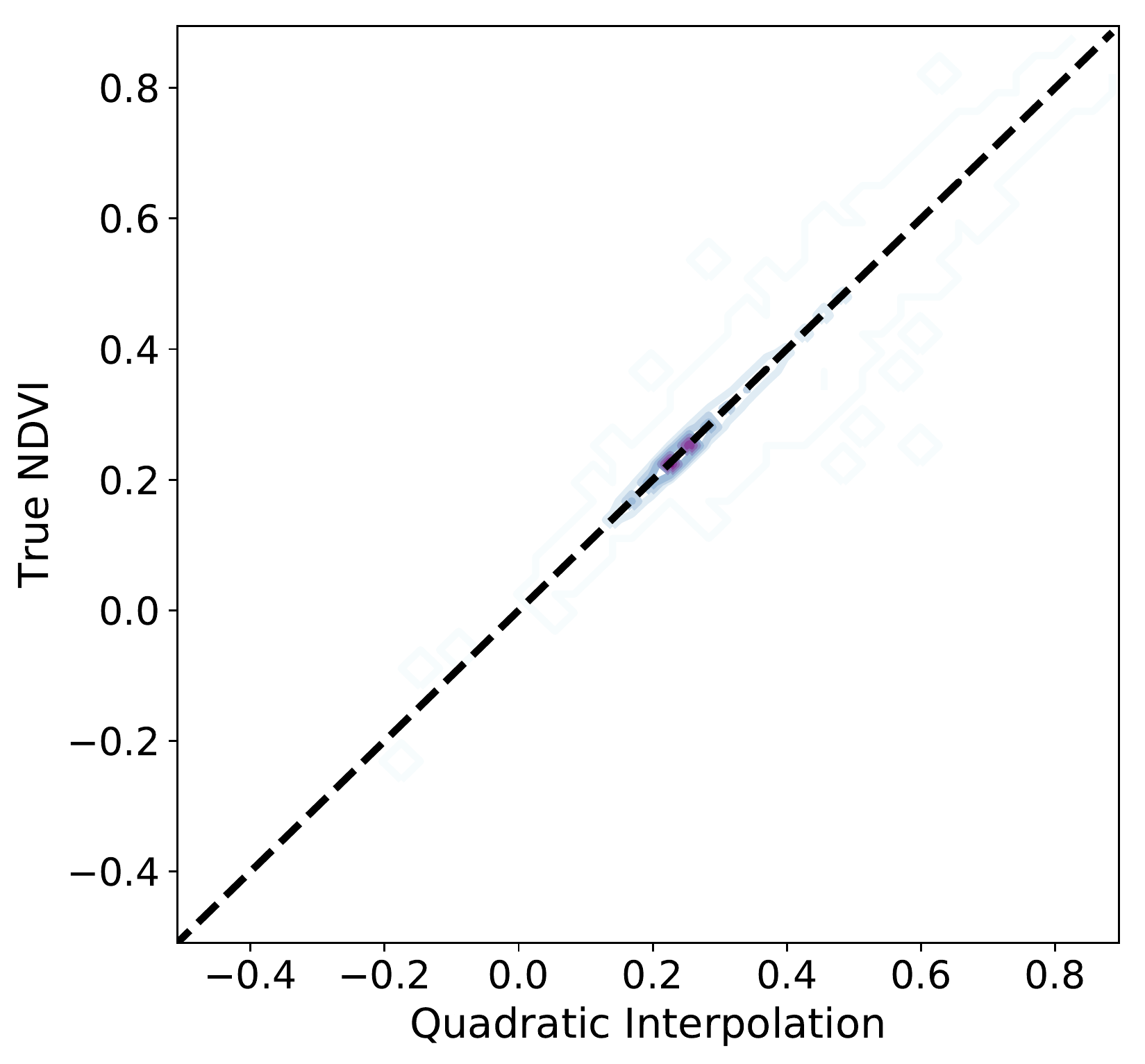} 
	\caption{Contour plot of MODIS observed and predicted NDVI values from 2000 pixels for gap-filling by quadratic interpolation.} \label{fig:interpScatter}
\end{figure}

For the MODIS data, GP, linear interpolation and quadratic interpolation all performed similarly well. Quadratic interpolation had the highest $R^2$-score, hence this method was chosen for gap-filling on the MODIS data. The higher interpolation $R^2$-scores for MODIS, compared to Landsat, imply that the MODIS data is less noisy than the Landsat data. Assuming that observations from MODIS and Landsat have similar signal-to-noise ratio, this can be explained by the higher temporal resolution of MODIS, and the compositing of multiple observations for the weekly gridded MODIS data. Fig.~\ref{fig:interpScatter} shows a contour plot of the true versus interpolated NDVI observations using the quadratic interpolation method. This again demonstrates that the interpolation doesn't introduce biases- the slope and intercept are approximately 1 and 0 respectively.

\begin{table}
	\caption{Comparison of interpolation methods as candidates for gap-filling on MODIS data.} 
	 \label{tab:comp_int2}
	\centering
	\begin{tabular}{lc} 
		\toprule
		\textbf{Method}  & \textbf{$R^2$-score} \\
		\midrule
		GP & 0.92 \\
		Linear & 0.93\\
		Quadratic & 0.94\\
		Cubic & 0.92\\
		Last value & 0.70\\
		Mean value & -0.02\\
		\bottomrule
	\end{tabular}
\end{table}
\vspace{2cm}

\newpage

\section{Further forecast results} \label{sec:further_results}

\begin{figure}
	\centering
	\includegraphics[trim = 30mm 0mm 0mm 10mm,width=5.4 cm]{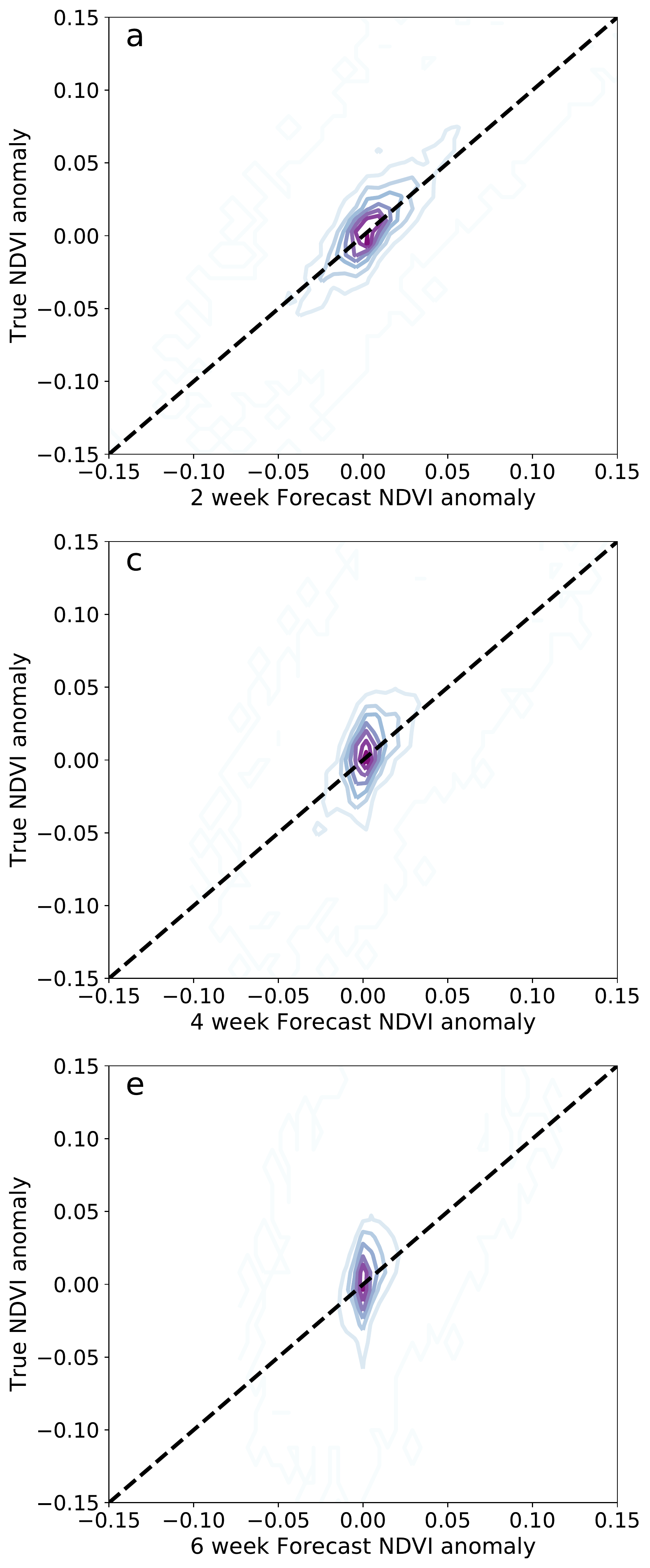} \qquad \includegraphics[trim = 21mm 0mm 0mm 0mm,width=5.4 cm]{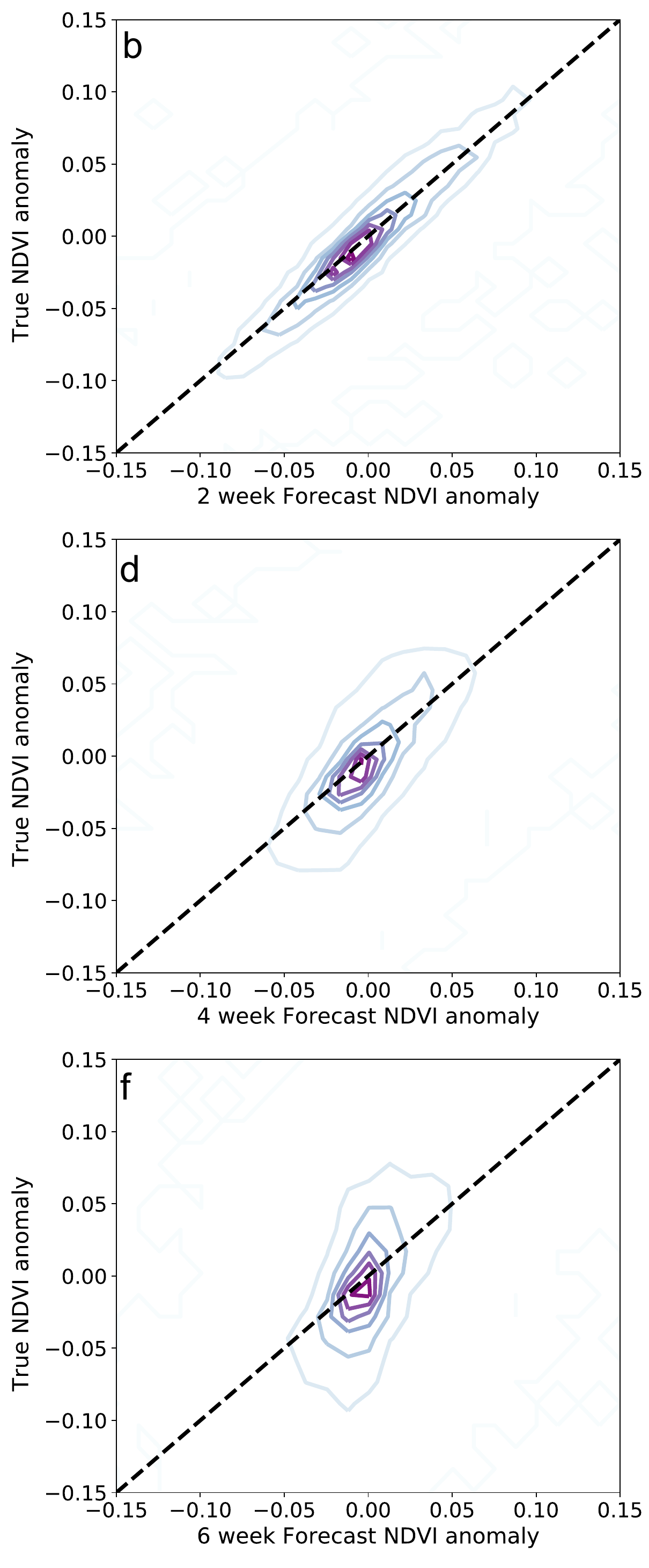}
	\caption{Contour plots of NDVI anomaly against two, four and six weeks NDVI anomaly forecasts. (a,c,e) show forecast performance for the GP method on Landsat data, and (b,d,f) show  forecast performance for the AR method on MODIS data (across the 19 regions for which a 4 week forecast was possible more than 50\% of the time, see main text for details).} \label{fig:contourNDVI}
\end{figure}

\begin{figure}[H]
	\centering
	\includegraphics[trim = 50mm 0mm 0mm 0mm,width=14 cm]{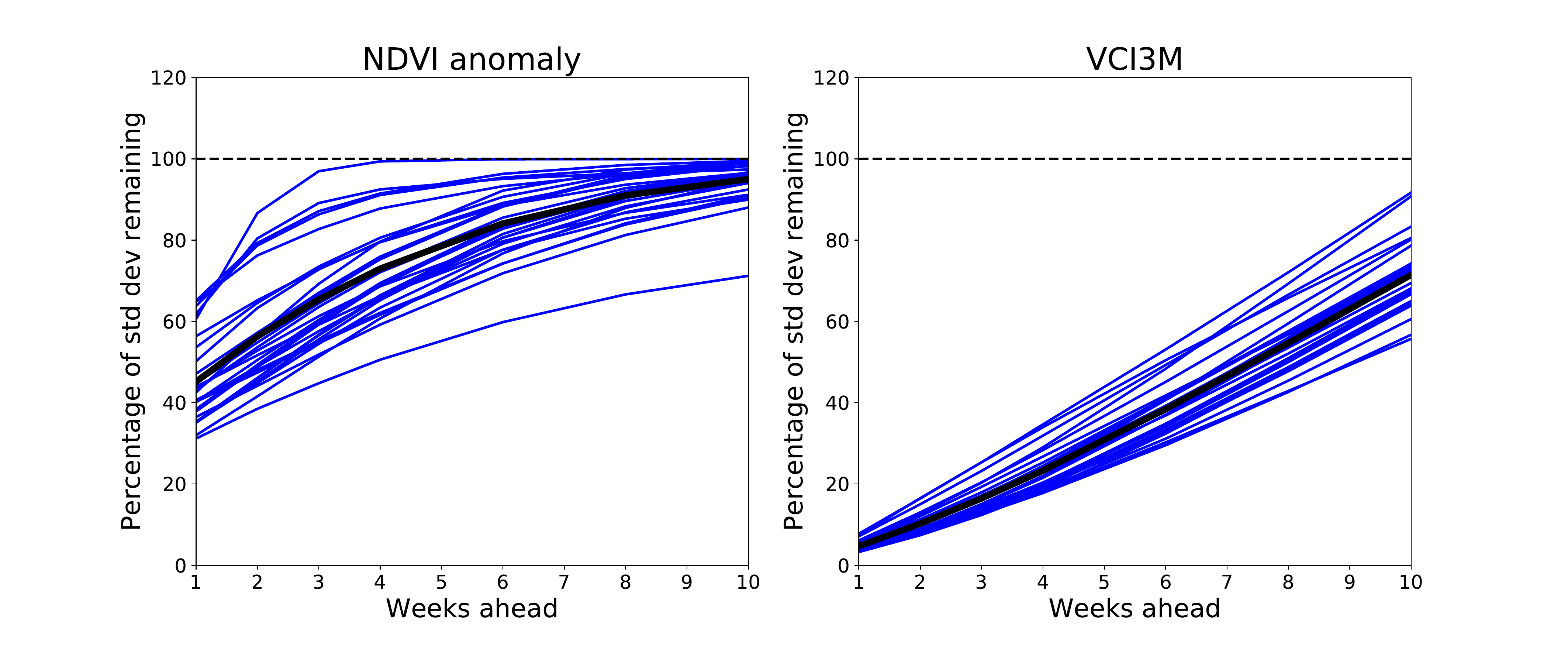} 
	\caption{Forecast performance with a lead time of 1 to 10 weeks using the GP method on the Landsat data, as given by percentage standard deviation remaining $S$, for (Left) NDVI anomaly, and (Right) VCI3M. The blue lines show results for the individual regions, and the black line shows the median across all regions.} \label{fig:GP_NDVI_forecast}
\end{figure}

\begin{figure}[H]
	\centering
	\includegraphics[trim = 20mm 0mm 0mm 0mm,width=12.5cm]{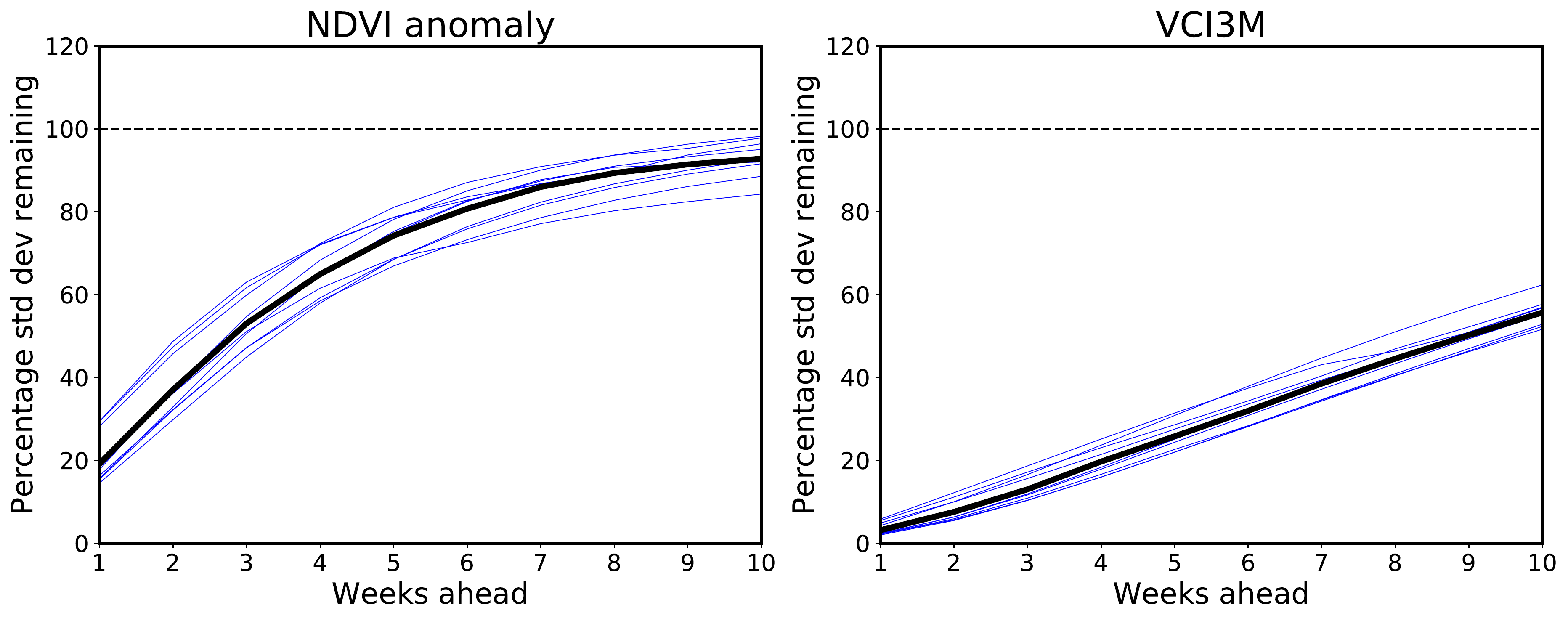} 
	\caption{Forecast performance with a lead time of 1 to 10 weeks using the AR method on the MODIS data, as given by percentage standard deviation remaining, for (Left) NDVI anomaly, and (Right) VCI3M. The blue lines show results for the individual regions for which a forecast is possible more than 50\% of the time, and the black line shows the median across all 19 of these regions.} \label{fig:NDVI_forecast}
\end{figure}

\begin{figure}[H]
	\centering
	\includegraphics[trim = 20mm 0mm 0mm 0mm,width=12.5cm]{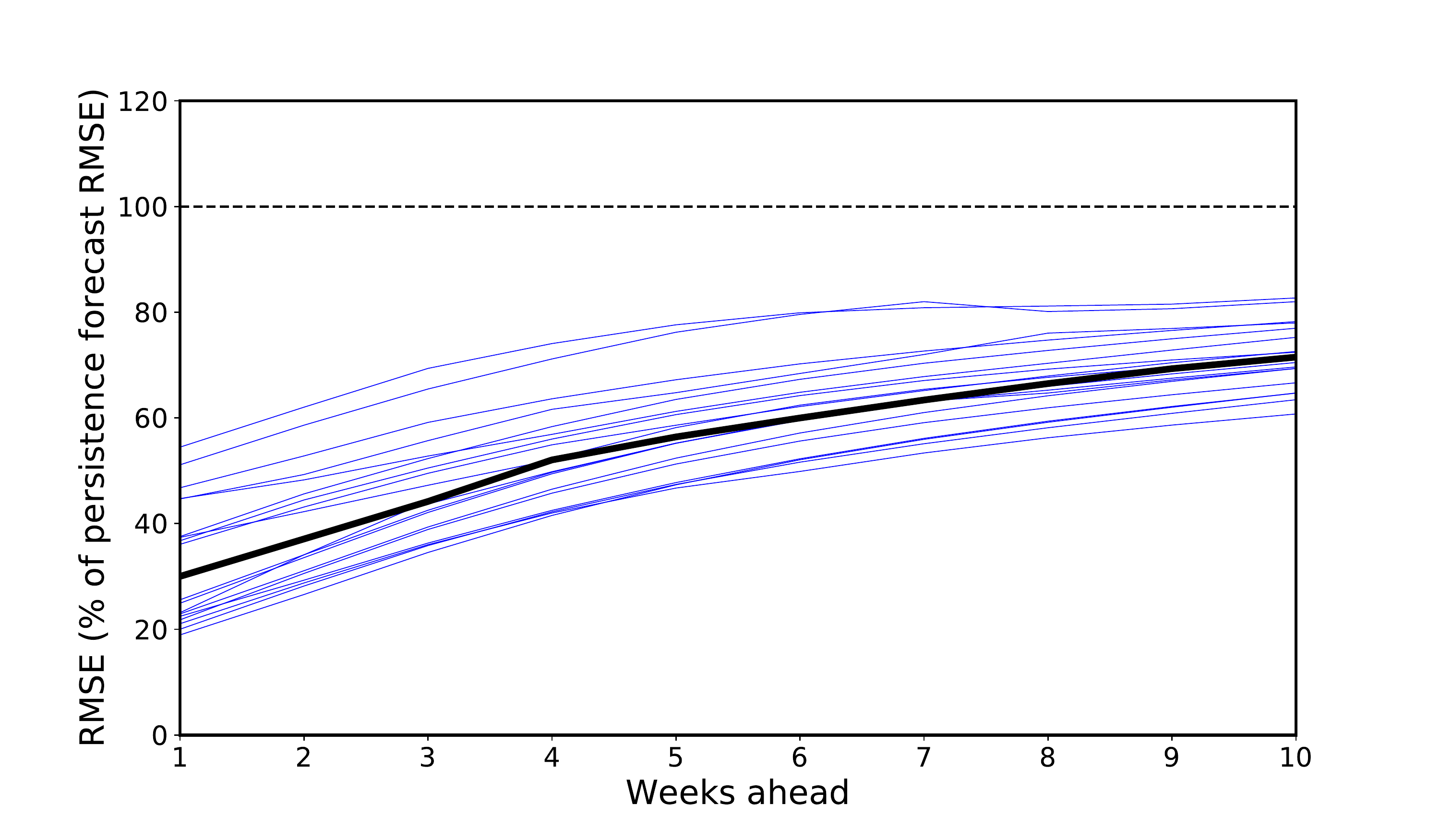} 
	\caption{Comparison of AR forecast with persistence forecast on the MODIS data. For lead times of 1 to 10 weeks, the RMSE of the AR forecast as a percentage of the RMSE of the persistence forecast. The blue lines show results for the individual regions for which a 4 week forecast is possible more than 50\% of the time, and the black line shows the median across these regions.} \label{fig:persistence}
\end{figure}

\begin{figure}[H]
	\centering
	\includegraphics[trim = 30mm 35mm 0mm 10mm,width=10 cm]{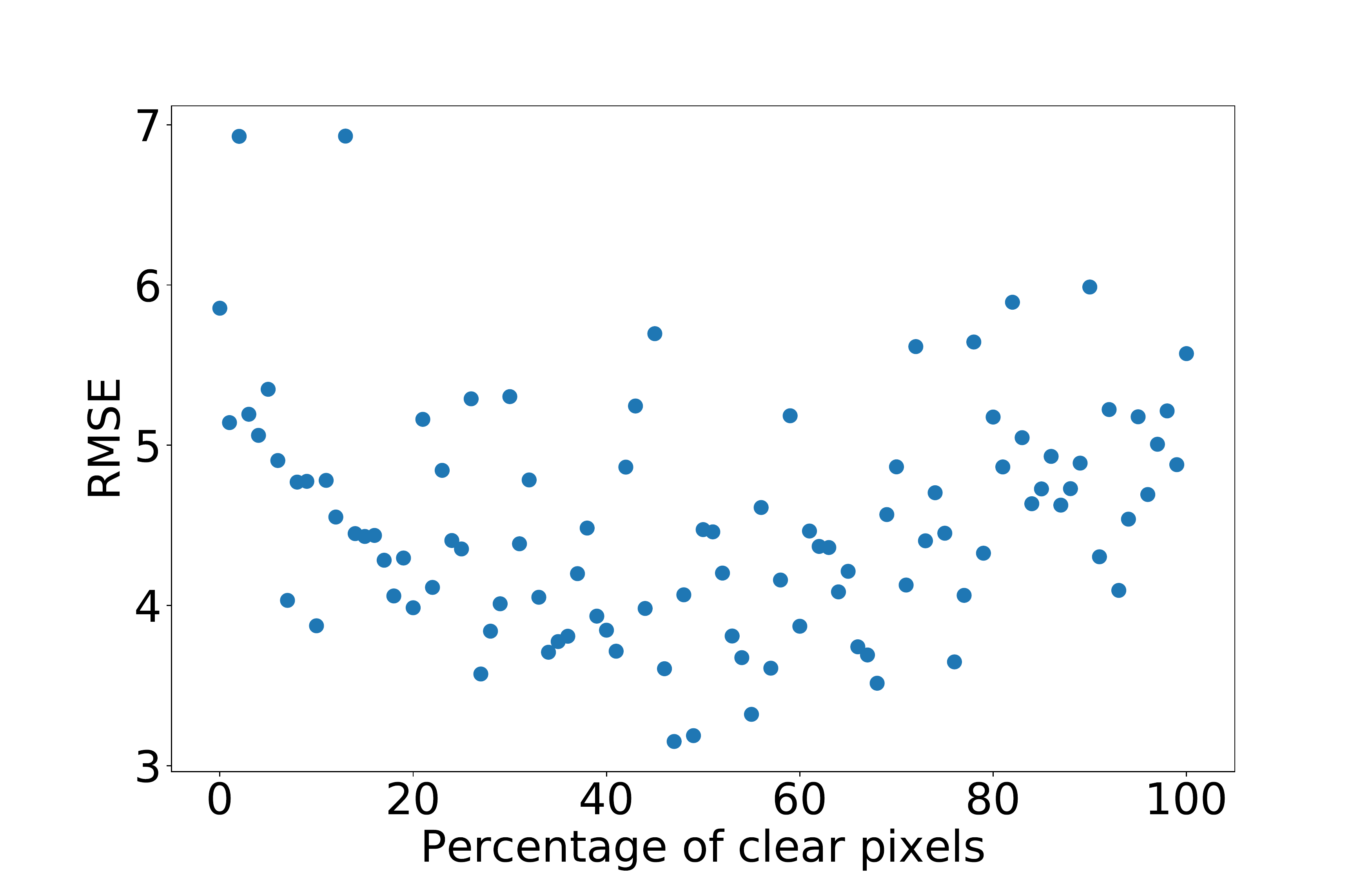} 
	\vspace{0.8cm}
	\caption{RMSE of 4 week forecast against percentage of clear pixels at most recent observation, for the AR method on the MODIS data. Plotted points are RMSE for each integer percentage of clear pixels. The Pearson correlation here is 0.01.} \label{fig:percentclear}
\end{figure}

\begin{figure} 
	\centering
	\includegraphics[trim = 0mm 0mm 5mm 20mm,width=12.3 cm]{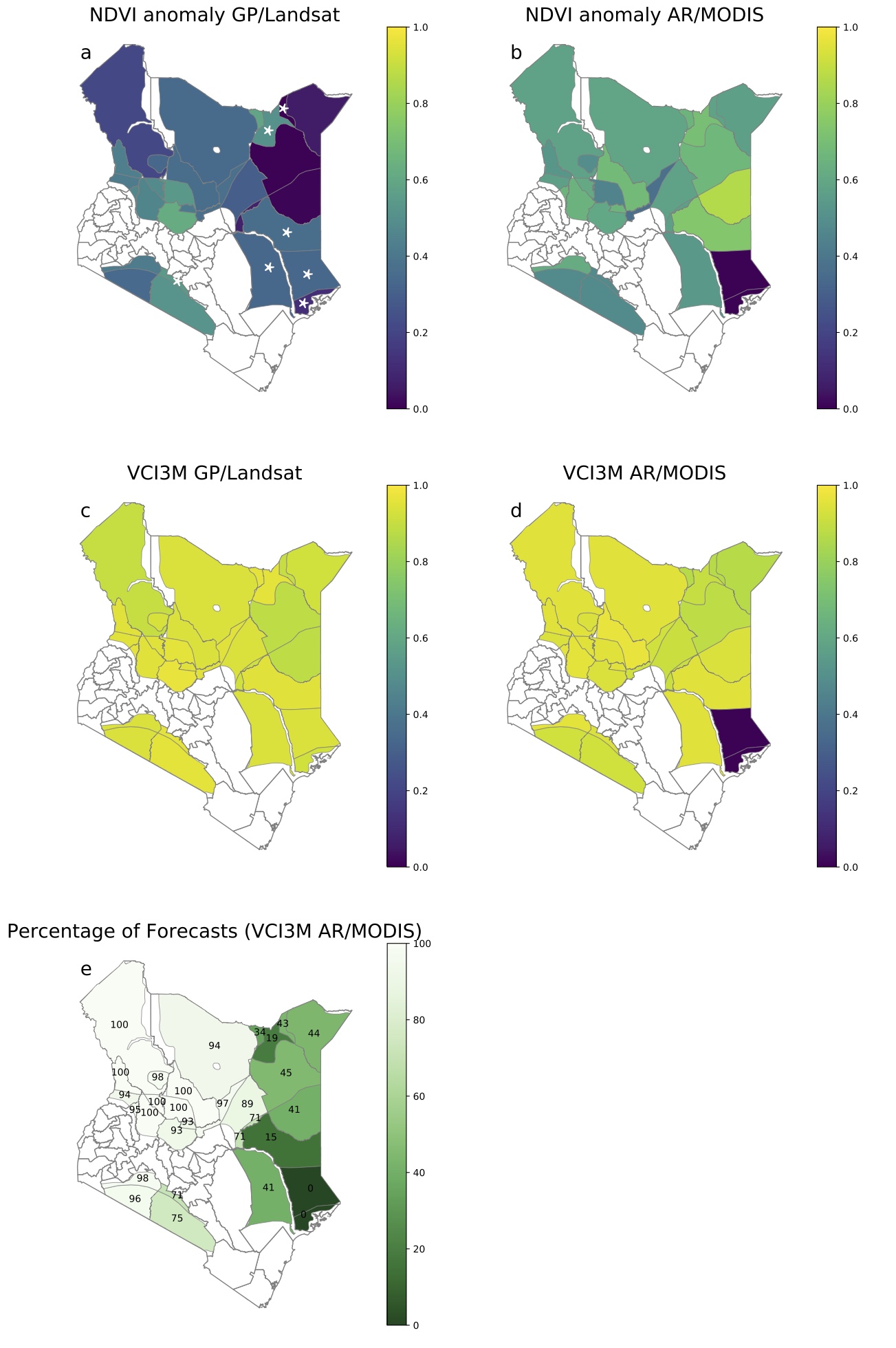}
	\vspace{0cm}
	\caption{Maps of NDVI anomaly and VCI3M 4 week forecast performance region-by-region for: (a) NDVI anomaly with GP method on Landsat data; (b) NDVI anomaly with AR method on MODIS data; (c) VCI3M with GP method on Landsat data; (d) VCI3M with AR method on MODIS data. In (a), asterisks indicate regions where selected pixels had a minimum of 180, rather than 250, clean observations. (e) shows the percentages of weeks that the AR method provided a 4 week VCI3M forecast.} \label{fig:R2plots}
\end{figure}

\begin{figure}[H]
	\centering
\includegraphics[trim = 30mm 35mm 0mm 10mm,width=5.4cm cm]{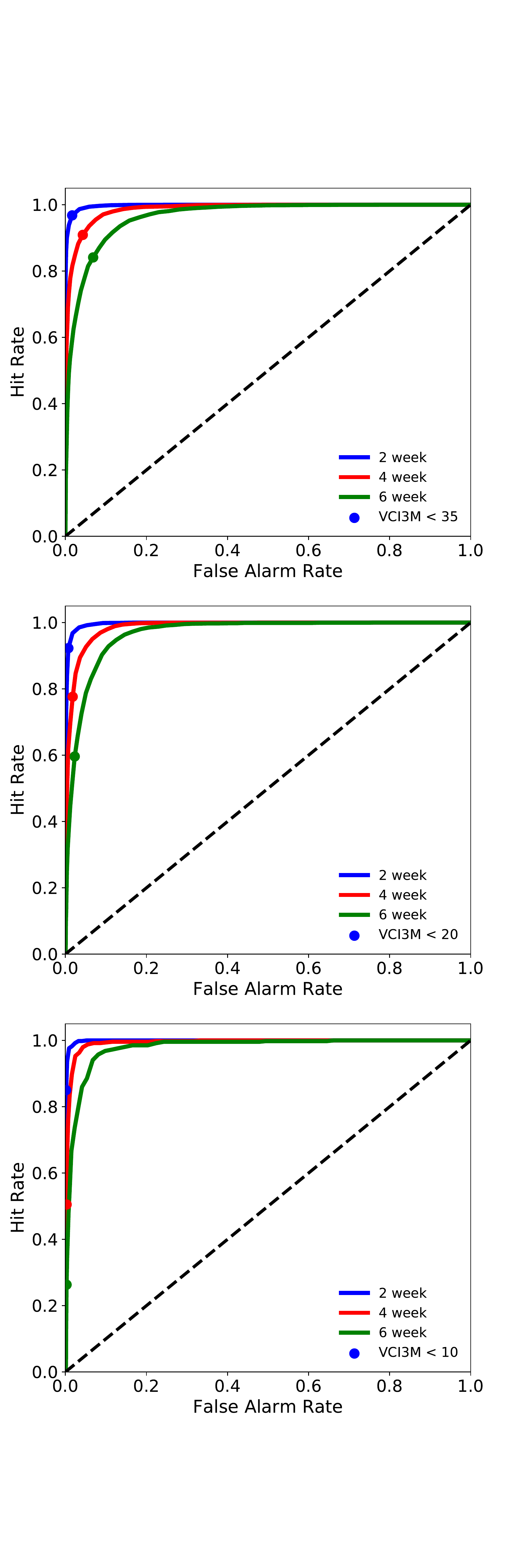} 
	\caption{ROC curves for predicting drought with drought defined at various NDMA thresholds. Possible hit rates against possible false alarm rates for the AR method on the MODIS data for the detection of: (Top) Any drought, VCI3M$<$35, (Middle) Severe or extreme drought VCI3M$<$20, (Bottom) Extreme drought} VCI3M$<$10. \label{fig:ROCotherdrought}
\end{figure}

Fig.~\ref{fig:contourNDVI} shows contour plots of forecast against actual NDVI anomaly data for the two methods, and Table \ref{tab:stats2} shows the $R^2$-scores, RMSE, slope and intercept from each of these plots.
Figs.~\ref{fig:GP_NDVI_forecast} and \ref{fig:NDVI_forecast} plot the forecast performance of the two methods in terms of percentage of standard deviation remaining $S$, for lead times of 1 to 10 weeks. For NDVI anomaly, for both methods, $S$ approaches the baseline of 100 as the lead time approaches 10 weeks, while for VCI3M, some forecast skill is still apparent at a lead time of 10 weeks. Fig.~\ref{fig:persistence} compares the performance of the AR VCI3M forecast with that of the persistence VCI3M forecast, on the MODIS data; the persistence forecast being simply the most recent observation. The AR forecast performs substantially better than the persistence forecast, for example, achieving a RMSE of approximately half that of the persistence forecast for a lead time of 4 weeks. The GP VCI3M forecast on the Landsat data achieves a similar improvement on the persistence forecast. Fig.~\ref{fig:percentclear} shows, for the MODIS/AR method, the average RMSE of a 4 week forecast against the percentage of pixels from which there was a clear observation during the week the forecast was made. Fig.~\ref{fig:R2plots} shows forecast performance region by region. Fig.~\ref{fig:ROCotherdrought} shows alternative ROC curves for drought prediction using the AR method on the MODIS data, based on different thresholds for defining drought.

\begin{table}
	\small
	\caption{Performance statistics of NDVI anomaly forecasts with lead times of 2, 4 and 6 weeks. Data for slope and intercept show ordinary least squares estimates $\pm$ standard error.} \label{tab:stats2}
	\centering
	\begin{tabular}{l|ccc|ccc} 
		\toprule
		& & \textbf{Landsat GP} & & &\textbf{MODIS AR} \\
		& \textbf{2} & \textbf{4} & \textbf{6} & \textbf{2} & \textbf{4} & \textbf{6} \\
		& \multicolumn{3}{c|}{\textbf{weeks }}& \multicolumn{3}{c}{\textbf{weeks }}\\

		\midrule
		$R^2$-score  &0.69  &0.46  &0.27  &0.85& 0.55& 0.33\\
		RMSE &0.029  &0.039  &0.045  &0.025& 0.043& 0.053\\
		slope &1.1$\pm0.0$  &1.2$\pm0.0$ &1.4$\pm0.0$ &1.0$\pm0.0$&1.0$\pm0.0$&1.0$\pm0.0$\\
		intercept &0.0$\pm0.0$ &0.0$\pm0.0$ &0.0$\pm0.0$&0.0$\pm0.0$&0.0$\pm0.0$&0.0$\pm0.0$\\
		
		\bottomrule
	\end{tabular}
\end{table}

\newpage

\subsection{Effect of including observations from other regions in the AR model} 

For the MODIS data, we tested to see whether we could improve the prediction of VCI3M by incorporating the past of VCI3M from a distinct region in the AR model, i.e.~Granger causality analysis was performed. Taking $X$ as the VCI3M of the region to be forecast, as in equation \eqref{eq:AR1}, and $Y$ to be the VCI3M from another region, the extended model was fit:
\begin{equation}
X_{t+n}=\sum_{i=0}^{p-1}a_iX_{t-i}+\sum_{i=0}^{q-1}b_iY_{t-i}+\epsilon^{\prime}_t\,,
\end{equation}
and Granger causality measured as the percentage reduction in RMSE obtained when the extended model is used instead of the previous (reduced) model \eqref{eq:AR1}. 

\begin{figure} 
	\centering
	\includegraphics[trim = 20mm 0mm 0mm 0mm,width=11 cm]{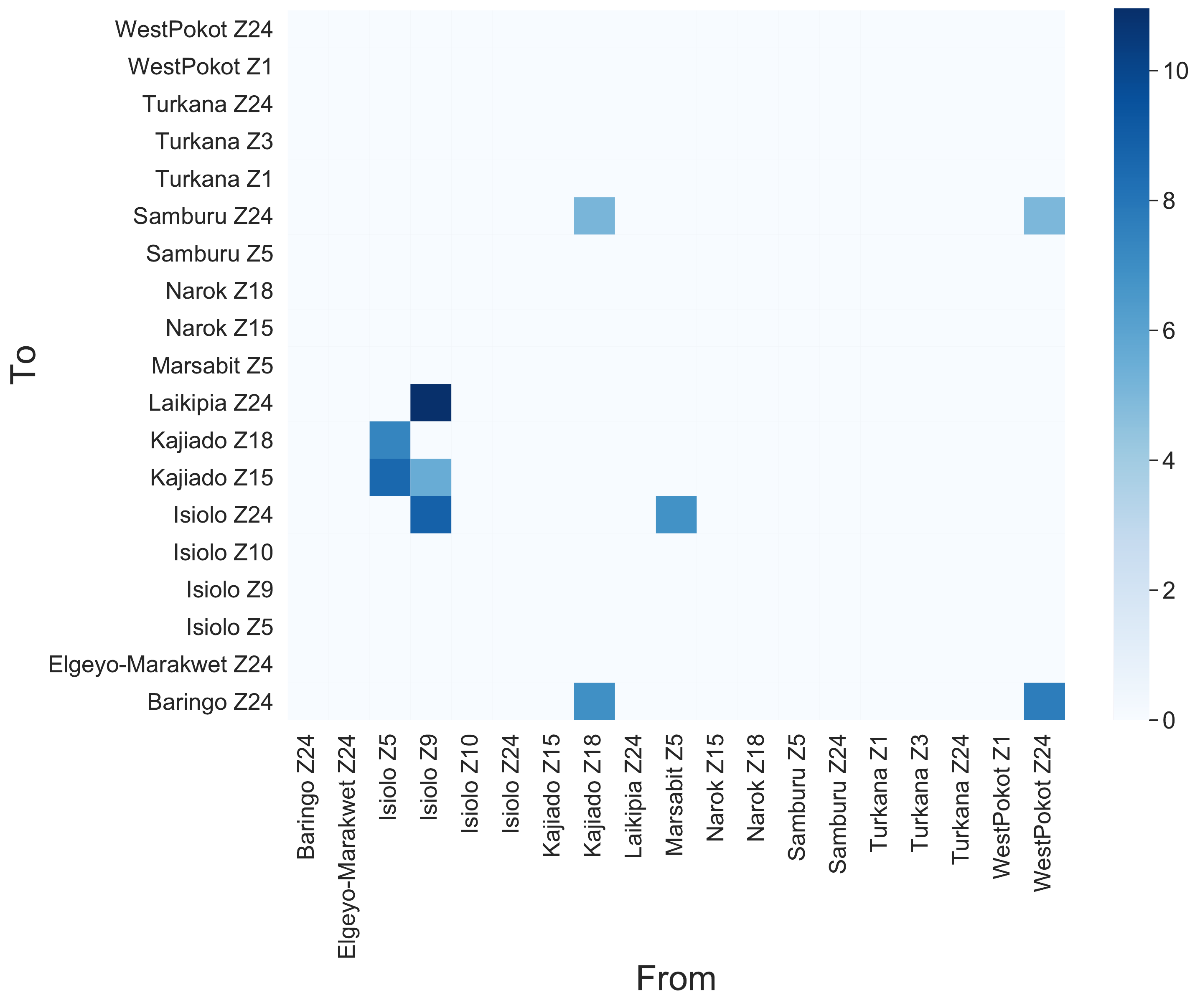}
	\caption{Granger causality of VCI3M} from each region to each other region, computed on the MODIS data, measured as percentage reduction in RMSE when observations from region `From' are added to the AR model for forecasting region `To' at a lead time of 4 weeks. Only substantial Granger causalities are shown, i.e.~those with percentage reduction in RMSE of more than $5\%$. \label{fig:GCheatmap}
\end{figure}

We tested for Granger causality of VCI3M from each region to each other region (within the set of regions for which predictions could be made more than 50\% of the time). That is, for each pair of distinct regions, $i$ and $j$, the 3 most recent observations from region $j$ were added to the AR forecast model for region $i$, and the RMSE was compared with that obtained without including observations from region $j$. There was not strong Granger causality of VCI3M between most regions. For only a few combinations was there a reduction in RMSE of more than $5\%$, see Fig.~\ref{fig:GCheatmap}. Nevertheless, these results suggest that, to create the optimal linear regression based forecasting method, data from all regions should be used. Future work will explore how best to extract any useful information from regions other than the one being forecast.

\end{document}